\documentclass[seceq,preprint]{ptptex}

\newcommand{\dz}{\frac{dz}{2\pi i}}
\newcommand{\norm}[1]{{}^\times_\times{#1}^\times_\times}
\newcommand{\onorm}[1]{{}^\otimes_\otimes{#1}^\otimes_\otimes}
\preprintnumber[3cm]{%<-- [..]: optional width of preprint # column.
hep-th/0406095\\YITP-04-31\\
}
\title{%        %You can use \\ for explicit line-break
Hybrid Superstring on $AdS_3$ \\
and \\
Space-Time Superconformal Symmetry
}

\author{%       %Use \scshape  for the family name
Hiroshi \textsc{Kunitomo}%
}

\inst{%         %Affiliation, neglected when [addenda] or [errata]
Yukawa Institute for Theoretical Physics,\\
Kyoto University, Kyoto 606-8502, Japan
}

\recdate{June 11, 2004}%            %Editorial Office will fill in this.

\abst{%         %this abstract is neglected when [addenda] or [errata]
Using a hybrid formalism, superstrings on $AdS_3\times S^1$ 
are studied in a manifestly supersymmetric manner. The world-sheet fields
in this description including superspace coordinates are obtained 
through a field redefinition from the corresponding ones in the RNS formalism.
The physical states are defined with the BRST cohomology as an $N=4$
topological string theory.
The physical spectra for some lower mass levels are investigated.
We identify two series of the space-time chiral primaries
which have already been obtained from the analysis 
in the RNS formalism. We find that they are described by the chiral 
and the vector supermultiplets, but the on-shell physical structure of
the latter depends on whether it is massive or massless.
The space-time supersymmetry on this background
is extended to the boundary $N=2$ superconformal symmetry.
We explicitly construct its generators and study how they act
on specific supermultiplets. 
}

\begin{document}

\maketitle

\section{Introduction}

The world-sheet symmetries play an important role during the
first evolutions in the superstring theory.
These symmetries are useful for studying the string
as a two-dimensional field theory, which has led to important
developments in the understanding of its perturbative behavior. 
From this viewpoint, the Ramond-Neveu-Schwarz (RNS) formalism 
is the most convenient formulation to obtain
space-time covariant results. The space-time supersymmetry,
on the other hand, is not manifest in this formalism
but, rather, is obtained by imposing the GSO projection on 
representations of the (world-sheet) $N=2$ superconformal symmetry.

In the next stage, however, the space-time symmetries
become more important for studying non-perturbative aspects 
of superstrings. The space-time supersymmetry, in particular,
becomes more important as a way to classify several BPS states 
belonging to some short representations of the supersymmetry. 
The spectra of these states are kept from changing due to 
the supersymmetry and can be treated independently of the 
string coupling constant.
The Green-Schwarz (GS) formalism has manifest space-time 
supersymmetry, and therefore it is the most convenient way to
investigate these states. Nevertheless, it is not known how to
quantize the GS superstrings while keeping all the space-time
supersymmetries manifest.

The formalism proposed and developed in Refs. \citen{B,BV,BVW} is 
a {\it hybrid} of these two, the RNS and the GS formalisms, 
and can be quantized in such a manner that preserves the manifest 
$D$-dimensional supersymmetry for $D<10$. The $D$-dimensional part 
is described by space-time superspace coordinates
(with conjugates of fermionic coordinates) and some additional bosons.
The remaining $(10-D)$-dimensional part, interpreted as representing 
some compactified space, is generally described by appropriate 
representations of the $N=2$ superconformal field theory. 
The physical states are defined with the BRST cohomology as 
constituting an $N=4$ topological string.\cite{BV} 
This formalism has already been applied to the study of superstrings 
on a variety of backgrounds, and its validity 
has been demonstrated.\cite{BVW,backgrounds,K}

In this paper, we apply the hybrid formalism to 
superstrings on $AdS_3\times S^1\times \mathcal{N}/U(1)$
with the manifest space-time (anti-de Sitter) supersymmetry 
on $AdS_3\times S^1$. 
These backgrounds have received much 
attention\cite{GKS}\tocite{HHS} as the simplest solvable 
model for studying AdS/CFT-duality\cite{adscft} beyond 
the supergravity approximation. Strings propagating on $AdS_3$, 
with an NS $B$-field background, are described by the $SL(2,R)$ 
WZW model with level $k$\cite{AdS3,AdS3-2} and are exactly solvable 
as a conformal field theory. Their spectrum has been studied 
in detail, and it has been found that it must include spectrally flowed 
representations.\cite{MO} These representations can be naturally 
incorporated by using a free-field realization as the discrete 
light-cone Liouville theory,\cite{HHS} in which the discrete 
light-cone momentum is identified with the spectral flow parameter.
Applying this framework to the RNS superstrings, 
two series of the space-time chiral primaries were found.\cite{HHS} 

From these world-sheet fields in the RNS formalism, 
we can obtain those in the hybrid formalism through 
a field redefinition. The four-dimensional sector describing 
$AdS_3\times S^1$ is represented by the superspace coordinates 
$(X^0,X^1,\phi_L,\widehat Y; \Theta^\pm,\bar\Theta^\pm)$,
the conjugates of the fermionic coordinates $(\mathcal{P}_\pm,
\bar{\mathcal{P}}_\pm)$, and an additional boson $\rho$.
The compactified space $\mathcal{N}/U(1)$ sector 
is characterized by the $N=2$ superconformal symmetry, generated by 
$(T_{\mathcal{N}/U(1)},G^\pm_{\mathcal{N}/U(1)},I_{\mathcal{N}/U(1)})$,
with central charge $c=9-6/k$.

The physical states in this formalism are defined with the BRST cohomology
as an $N=4$ topological string theory.\cite{BV} 
We explicitly identify physical spectra corresponding to
the two series of the space-time chiral primaries which have been
obtained in the RNS formalism.\cite{HHS}
The supermultiplet structures of these series are also clarified.
The first series is described by the chiral supermultiplet including
a scalar, a two-component spinor and an auxiliary field. 
In the second series, the two cases in which the $\mathcal{N}/U(1)$ sector 
is excited and not excited must be distinguished.
We find that the former (latter)
is given by the massive (massless) vector supermultiplet.
The space-time supersymmetry on this background $AdS_3\times S^1$
can be extended to the boundary $N=2$ superconformal symmetry 
with the central charge $c=6kp$.
This is closed off shell in the hybrid formalism.
We explicitly construct its  generators and study how they 
act on the specific fields in the supermultiplet. 
Because the spectral flow operation does not commute 
with the world-sheet Hamiltonian, 
the generators act as spectrum generating operators on generic 
states with non-vanishing light-cone momentum $p\ne0$. 
Only two of the supersymmetries are closed on a supermultiplet,
while the others generate new physical states with different masses.
On the special states with vanishing light-cone 
momentum $p=0$, by contrast, all of the infinite 
supersymmetries are realized on a supermultiplet.

This paper is organized as follows. 
We begin with a brief review of the RNS superstring on 
$AdS_3\times S^1$ in \S\ref{RNS}. The (super) WZW model 
on this background is described in terms of a free-field
realization as the discrete light-cone Liouville theory.\cite{HHS} 
We reformulate it as an $N=4$ topological string theory 
for later convenience. In \S\ref{hybrid}, hybrid world-sheet 
fields are introduced through a field redefinition.
The model can be completely rewritten in terms of 
these hybrid fields. This makes all the space-time 
supersymmetries manifest. The generators of the boundary 
$N=2$ superconformal algebra are constructed explicitly.
In \S\ref{flow} we construct the Hilbert space of the hybrid
superstring by extending the spectral flow operation to 
the currents consisting of the space-time spinor fields.
The physical spectrum for some lower mass levels is 
investigated in \S\ref{physspec}. They are identified
with the two series of the space-time chiral primaries
and represented by the chiral and the vector supermultiplets,
respectively. We study, in \S\ref{stsusy}, 
how the off-shell supersymmetries are realized on these supermultiplets.
Section \ref{concl} is devoted to summary and discussion. 
Three appendices are added for some useful details.
In Appendix A, we rewrite the space-time $N=2$ superconformal 
generators using the currents of the global $sl(1|2)$ 
symmetry.\cite{Ito} 
The hybrid fields introduced in this paper are
analogs of the {\it chiral} coordinates 
$(y^m,\theta^\alpha,\bar{\theta}^\alpha)$ in flat four-dimensional
space-time, which is convenient for studying the chiral supermultiplet.\cite{WB}
We give a similarity transformation to obtain
the {\it real} coordinates with proper hermiticity in Appendix B.
The space-time and the world-sheet superconformal generators are
transformed. We show, in Appendix C, how the fields are transformed 
under the bosonic symmetry of the $N=2$ superconformal symmetry.
These transformation laws are needed to confirm that the symmetries 
actually satisfy the $N=2$ superconformal algebra.

\section{RNS superstrings on $AdS_3\times S^1\times\mathcal{N}/U(1)$}\label{RNS}

Let us start by briefly reviewing how superstrings propagating on
$AdS_3\times S^1\times\mathcal{N}/U(1)$ are described in the RNS
formalism. It is given by the tensor 
product of the $N=2$ superconformal field theories representing
each component space.

The $AdS_3$ sector is described by a world-sheet
supersymmetric extension of the $sl(2)$ current algebra,\footnote{
We consider only the holomorphic sector in this paper.
It is easy to combine it with the anti-holomorphic sector 
if necessary.\cite{GKS}\tocite{HHS}}
\begin{align}
  j^3(z)j^3(w)&\sim-\frac{(k+2)/2}{(z-w)^2},\nonumber\\
j^3(z)j^{\pm\pm}(w)&\sim\frac{\pm j^{\pm\pm}(w)}{z-w},\nonumber\\
j^{++}(z)j^{--}(w)&\sim\frac{k+2}{(z-w)^2}-\frac{2j^3(w)}{z-w},\label{sl2alg}
\end{align}
by introducing three free fermions,
\begin{equation}
  \psi^+(z)\psi^-(w)\sim\frac{2}{z-w},\qquad
 \psi^3(z)\psi^3(w)\sim-\frac{1}{z-w}.
\end{equation}
The level of this algebra is taken to be such that
the algebra of the total currents,
\begin{align}
 J^{\pm\pm}=&j^{\pm\pm}\pm\psi^\pm\psi^3,\nonumber\\
 J^3=&j^3+\frac{1}{2}\psi^+\psi^-,
\end{align}
is of level $k$.
For the bosonic currents (\ref{sl2alg}), 
we use the free-field realization\cite{sl2free,HHS}
\begin{align}
j^{\pm\pm}=&e^{\mp\beta i(X^0+X^1)}\left(
-\frac{1}{\beta}i\partial X^1\pm\frac{1}{Q}\partial\phi_L\right),\nonumber\\
j^3 =&\frac{1}{\beta}i\partial X^0,\label{bsl2}
\end{align}
where $\beta=\sqrt{\frac{2}{k+2}}$ and $Q=\sqrt{\frac{2}{k}}$.

The $S^1$ sector is simply represented by a pair of free fields 
$(Y, \psi^4)$ satisfying
\begin{align}
 Y(z)Y(w)&\sim-\log(z-w),\nonumber\\
 \psi^4(z)\psi^4(w)&\sim\frac{1}{z-w}.
\end{align}
We define a $U(1)$ current as
\begin{align}
 J^Y=&\frac{2}{Q}i\partial Y,
\end{align}
for later use.

This description of the RNS superstring on $AdS_3\times S^1$ possesses 
the $N=2$ world-sheet superconformal symmetry generated by
\begin{align}
 T^{(4)}=&-\frac{1}{2}\partial X^\mu\partial X_\mu
-\frac{1}{2}\partial\phi_L\partial\phi_L-\frac{Q}{2}\partial^2\phi_L
-\frac{1}{2}\partial Y\partial Y\nonumber\\
&\hspace{1cm}
-\frac{1}{4}\psi^+\partial\psi^-
-\frac{1}{4}\psi^-\partial\psi^+
+\frac{1}{2}\psi^3\partial\psi^3
-\frac{1}{2}\psi^4\partial\psi^4,\nonumber\\
%%%%%%%%%%%%%%%%%%%%%%%%%%%%%%%%
G^{\pm(4)}=&\frac{Q}{2}\psi^\pm j^{\mp\mp}
+\frac{1}{2}(\psi^4\mp\psi^3)\left(
i\partial Y\pm Qj^3\right)
\pm\frac{Q}{4}(\psi^4\mp\psi^3)\psi^+\psi^-,\nonumber\\
%%%%%%%%%%%%%%%%%%%%%%%%%%%%%%%%
I^{(4)}=&Q^2j^3+
\frac{1}{2}\left(1+Q^2\right)\psi^+\psi^-
+\psi^4\psi^3,
\end{align}
with the central charge $c=6+6/k$.

For the compactified space $\mathcal{N}/U(1)$ sector,
we denote the $N=2$ superconformal generators as 
$(T_{\mathcal{N}/U(1)},G^\pm_{\mathcal{N}/U(1)},I_{\mathcal{N}/U(1)})$.
Their central charge has to be $c=9-6/k$ for criticality.
The string state is given by an arbitrary unitary (rational)
representation of this $N=2$ superconformal algebra characterized 
by two quantum numbers, $(\Delta_\mathcal{N},Q_\mathcal{N})$, 
the conformal weight and the $U(1)$ charge.

In addition to these {\it matter} fields, the superconformal ghosts
$(b,c)$ and $(\beta,\gamma)$ satisfying
\begin{equation}
 c(z)b(z)\sim\gamma(z)\beta(w)\ \sim\ \frac{1}{z-w}
\end{equation}
must be introduced to quantize the RNS superstring covariantly.
The ghost sector also possesses the $N=2$ superconformal symmetry generated by
\begin{align}
 T_{gh}=&-2b\partial c-\partial bc
-\frac{3}{2}\beta\partial\gamma-\frac{1}{2}\partial\beta\gamma,\nonumber\\
%%%%%%%%%%%
G^+_{gh}=&\frac{3}{2}\beta\partial c+\partial\beta c,\qquad
%%%%%%%%%%%
G^-_{gh}=-2b\gamma,\nonumber\\
%%%%%%%%%%%
I_{gh}=&2bc+3\beta\gamma,
\end{align}
although only its $N=1$ subset, $(T_{gh},G_{gh}=G^+_{gh}+G^-_{gh})$,
is familiar. The physical Hilbert space is defined by the cohomology 
\begin{equation}
\mathcal{H}_{{\rm phys}}={\rm Ker}Q_{{\rm BRST}}/{\rm Im}Q_{{\rm BRST}}
\label{cohomology}
\end{equation}
of the BRST charge
\begin{equation}
 Q_{{\rm BRST}}=\oint\dz\left(c\left(T_m+\frac{1}{2}T_{gh}\right)
+\gamma\left(G_m+\frac{1}{2}G_{gh}\right)\right),\label{brst}
\end{equation}
where
\begin{equation}
 T_m=T^{(4)}+T_{\mathcal{N}/U(1)},\qquad
G_m=G^{+(4)}+G^{-(4)}+G^+_{\mathcal{N}/U(1)}+G^-_{\mathcal{N}/U(1)}.
\end{equation}

In order to construct space-time supercharges,
we need to bosonize the world-sheet fermions and the $U(1)$
current $I_{\mathcal{N}/U(1)}$ as
\begin{subequations}
\begin{align}
 \psi^+\psi^-=&2i\partial H_0,\qquad
 \psi^3\psi^4=i\partial H_1,\label{psifb}\\
 I_{\mathcal{N}/U(1)}=&-\alpha i\partial H_2,\label{u1fb}
\end{align}
\end{subequations}
where $\alpha^2=3-Q^2$. The bosons 
$H_I(z)\ (I=0,1,2)$ satisfy the standard OPEs:
\begin{equation}
 H_I(z)H_J(w)\sim -\delta_{IJ}\log(z-w).
\end{equation}
The superconformal ghosts must be also bosonized as\cite{FMS}
\begin{align}
 c=&e^\sigma,\qquad b=e^{-\sigma},\nonumber\\
%%%%%%%%%%%%%%%%%%%%%%%%%%%
 \gamma=&\eta e^\phi=e^{\phi-\chi},\nonumber\\
%%%%%%%%%%%%%%%%%%%%%%%%%%%
 \beta=&e^{-\phi}\partial\xi=\partial\chi e^{-\phi+\chi},
\label{bg}
\end{align}
with
\begin{align}
 \phi(z)\phi(w)&\sim-\log(z-w),\nonumber\\
 \sigma(z)\sigma(w)&\sim\chi(z)\chi(w)\sim+\log(z-w).
\end{align}
Here, it is important to note that the Hilbert space of 
the original bosonic ghosts, $(\beta,\gamma)$, is different 
from that of the bosonized fields $(\phi,\xi,\eta)$,
or equivalently $(\phi,\chi)$, since the zero-mode, $\xi_0$, 
is not included in the bosonization formulas (\ref{bg}). 
The former (latter) is called the small 
(large) Hilbert space $\mathcal{H}_{{\rm small}}$ 
($\mathcal{H}_{{\rm large}}$). 
The BRST cohomology (\ref{cohomology}) is defined in
$\mathcal{H}_{{\rm small}}$. This extension of 
the Hilbert space is essential to realize
supersymmetry in the RNS formalism.

Now, four supercharges can be constructed in the familiar 
$-\frac{1}{2}$-picture as\cite{AdS3-2,HHS}
\begin{align}
  \mathcal{G}^{+(-\frac{1}{2})}_{\frac{1}{2}}=&
k^{\frac{1}{4}}\oint\dz e^{
-\frac{1}{2}\phi+\frac{i}{2}(H_0-H_1+\sqrt{3}H'_2)},\nonumber\\
%%%%%%%%%%%%%%%%%%%%%%%%%%%%%%%%%%%
 \mathcal{G}^{+(-\frac{1}{2})}_{-\frac{1}{2}}=&
k^{\frac{1}{4}}\oint\dz e^{
-\frac{1}{2}\phi+\frac{i}{2}(-H_0+H_1+\sqrt{3}H'_2)},\nonumber\\
%%%%%%%%%%%%%%%%%%%%%%%%%%%%%%%%%%%
 \mathcal{G}^{-(-\frac{1}{2})}_{\frac{1}{2}}=&
k^{\frac{1}{4}}\oint\dz e^{
-\frac{1}{2}\phi+\frac{i}{2}(H_0+H_1-\sqrt{3}H'_2)},\nonumber\\
%%%%%%%%%%%%%%%%%%%%%%%%%%%%%%%%%%%
 \mathcal{G}^{-(-\frac{1}{2})}_{-\frac{1}{2}}=&
k^{\frac{1}{4}}\oint\dz e^{
-\frac{1}{2}\phi+\frac{i}{2}(-H_0-H_1-\sqrt{3}H'_2)},\label{susyorg}
\end{align}
where $H'_2$ is the linearly transformed boson defined by
\begin{subequations}
\begin{align}
 \sqrt{3}H'_2=&\alpha H_2+QY,\label{u1mod}\\
 \sqrt{3}Y'=&-QH_2+\alpha Y.
\end{align}
\end{subequations}
The generators in (\ref{susyorg}), however, satisfy
the peculiar algebra\cite{GP,HHS}
\begin{alignat}{2}
  \left\{\mathcal{G}^+_{\pm\frac{1}{2}},\mathcal{G}^-_{\pm\frac{1}{2}}\right\}=&
  \frac{1}{Q}\oint\dz e^{-\phi}\psi^\pm,\ &\ 
 \left\{\mathcal{G}^+_{\pm\frac{1}{2}},\mathcal{G}^-_{\mp\frac{1}{2}}\right\}=&
  \frac{1}{Q}\oint\dz e^{-\phi}\left(\psi^3\mp\psi^4\right),\label{pecsl12}
\end{alignat}
which is equivalent to the familiar supersymmetry only for the on-shell
physical states. This can be improved by changing the picture for half 
of the supercharges $\mathcal{G}^{-(-\frac{1}{2})}_{\pm\frac{1}{2}}$ 
to $+\frac{1}{2}$, so that the right-hand sides of the algebra
(\ref{pecsl12}) become ghost independent, {\it i.e.} expressed in the $0$-picture:
\begin{align}
  \mathcal{G}^{-(\frac{1}{2})}_{\frac{1}{2}}=&
k^{\frac{1}{4}}\oint\dz\Bigg(
e^{\frac{3}{2}\phi+\frac{i}{2}(
H_0+H_1-\sqrt{3}H'_2)-\sigma-\chi}\nonumber\\
&\hspace{5mm}
+\frac{Q}{\sqrt{2}}
e^{\frac{1}{2}\phi-\frac{i}{2}(
H_0-H_1+\sqrt{3}H'_2)}
e^{-\beta i( X^0+ X^1)}
(\frac{1}{\beta}i\partial  X^1-\frac{1}{Q}\partial\phi_L)
\nonumber\\
&\hspace{10mm}
-\frac{Q}{\sqrt{2}}
e^{
\frac{1}{2}\phi-\frac{i}{2}(-H_0+H_1+\sqrt{3}H'_2)}
(\frac{1}{\beta}i\partial  X^0
+\frac{1}{Q}i\partial Y+i\partial H_0)
\nonumber\\
&\hspace{15mm}
+\frac{Q}{\sqrt{2}}e^{
\frac{1}{2}\phi+\frac{i}{2}(H_0+3H_1-\sqrt{3}H'_2)}
-e^{\frac{1}{2}\phi
+\frac{i}{2}(H_0+H_1-\sqrt{3}H'_2)}G^-_{\mathcal{N}/U(1)}
\Bigg),\nonumber\\
%%%%%%%%%%%%%%%%%%%%%%%%%%%%%
 \mathcal{G}^{-(\frac{1}{2})}_{-\frac{1}{2}}=&
k^{\frac{1}{4}}\oint\dz\Bigg(
e^{\frac{3}{2}\phi+\frac{i}{2}(
-H_0-H_1-\sqrt{3}H'_2)-\sigma-\chi}\nonumber\\
&\hspace{5mm}
+\frac{Q}{\sqrt{2}}e^{\frac{1}{2}\phi+\frac{i}{2}(
H_0-H_1-\sqrt{3}H'_2)}
e^{\beta i( X^0+ X^1)}
(\frac{1}{\beta}i\partial  X^1
+\frac{1}{Q}\partial\phi_L)
\nonumber\\
&\hspace{10mm}
-\frac{Q}{\sqrt{2}}
e^{\frac{1}{2}\phi+\frac{i}{2}(-H_0+H_1-\sqrt{3}H'_2)}
(\frac{1}{\beta}i\partial  X^0
-\frac{1}{Q}i\partial Y+i\partial H_0)
\nonumber\\
&\hspace{15mm}
-\frac{Q}{\sqrt{2}}e^{
\frac{1}{2}\phi+\frac{i}{2}(-H_0-3H_1-\sqrt{3}H'_2)}
+e^{\frac{1}{2}\phi
+\frac{i}{2}(-H_0-H_1-\sqrt{3}H'_2)}G^-_{\mathcal{N}/U(1)}
\Bigg).
\end{align}
We simply write 
$(\mathcal{G}^{+(-\frac{1}{2})}_{\pm\frac{1}{2}},
\mathcal{G}^{-(\frac{1}{2})}_{\pm\frac{1}{2}})=
(\mathcal{G}^+_{\pm\frac{1}{2}},
\mathcal{G}^-_{\pm\frac{1}{2}})$ hereafter
and do not change the picture further.
These picture-changed supercharges satisfy the familiar 
space-time ({\it AdS}) supersymmetry algebra $sl(1|2)$:
\begin{equation}
  \left\{\mathcal{G}^+_{\pm\frac{1}{2}},\mathcal{G}^-_{\pm\frac{1}{2}}\right\}=
  \mathcal{L}_{\pm1},\quad 
 \left\{\mathcal{G}^+_{\pm\frac{1}{2}},\mathcal{G}^-_{\mp\frac{1}{2}}\right\}=
  \mathcal{L}_0\pm\frac{1}{2}\mathcal{I}_0,
\end{equation}
where 
\begin{equation}
  \mathcal{L}_{\pm1}=-\oint\dz J^{\pm\pm},\quad 
%%%%%%%%%%%%%%%%%%%
 \mathcal{L}_0=-\oint\dz J^3,\quad
%%%%%%%%%%%%%%%%%%%
 \mathcal{I}_0=\oint\dz J^Y
\end{equation}
are the generators of the bosonic subgroup $SL(2)\times U(1)$.
This {\it global} $sl(1|2)$ symmetry can be extended to 
the infinite-dimensional $N=2$ superconformal symmetry,
and we explicitly construct its generators in \S\ref{hybrid}.

Before closing this section, we reconsider the physical state conditions
to develop the hybrid formalism in the next section.
As mentioned above, the physical states are originally defined with
the BRST cohomology in the small Hilbert space $\mathcal{H}_{{\rm small}}$. 
However, because $\mathcal{H}_{{\rm small}}$ is not sufficient to realize 
the space-time supersymmetry, we should modify them so as to obtain conditions 
in $\mathcal{H}_{{\rm large}}$ as
\begin{subequations}\label{pscorg}
\begin{align}
 Q_{{\rm BRST}}|\psi\rangle=&0,\nonumber\\
|\psi\rangle\sim&|\psi\rangle+\delta|\psi\rangle,\qquad
\delta|\psi\rangle\ =\ Q_{{\rm BRST}}|\Lambda\rangle,\nonumber\\
\eta_0|\psi\rangle=&\eta_0|\Lambda\rangle\ =\ 0,\label{psc}
\end{align}
where $|\psi\rangle, |\Lambda\rangle\in\mathcal{H}_{{\rm large}}$.
Additionally, we require that the physical states have ghost number $1$,
{\it i.e.},
\begin{equation}
  Q_{{\rm gh}}|\psi\rangle=|\psi\rangle,\label{gncond}
\end{equation}
\end{subequations}
counted by the charge\footnote{
This definition of the ghost number is related to 
the familiar one, $N_c=\oint\dz(cb-\gamma\beta)$, 
through the relation $Q_{gh}=N_c-\mathcal{R}$,
where $\mathcal{R}=\oint\dz(\xi\eta-\partial\phi)$ 
is the picture counting operator. 
The difference between the two definitions is a constant 
in a given picture.}
\begin{equation}
 Q_{{\rm gh}}=\oint\dz\left(cb-\xi\eta\right).
\end{equation}
The conditions given in (\ref{pscorg}) have a natural 
interpretation as an $N=4$ topological string theory,
as we now describe.

We note that there is the hidden twisted $N=4$ superconformal
symmetry in $\mathcal{H}_{{\rm large}}$ generated by\cite{BV}
\begin{align}
 T=&T_m+T_{{\rm gh}},\nonumber\\
%%%%%%%%%%%%%%%%%%%%%%%%%%%%%%
 G^+=&J_{{\rm BRST}},\nonumber\\
%%%%%%%%%%%%%%%%%%%%%%%%%%%%%%
 =& c\left(T_m+T_{\beta\gamma}\right)
+\gamma G_m-\gamma^2b+c\partial cb+\partial(c\xi\eta)+\partial^2c,
\nonumber\\
%%%%%%%%%%%%%%%%%%%%%%%%%%%%%%
G^-=&b,\qquad
 \widetilde G^+\ =\ \eta,\qquad
 \widetilde G^-\ =\ \xi T-b\left\{Q_{{\rm BRST}},\xi\right\}+\partial^2\xi,
\nonumber\\
%%%%%%%%%%%%%%%%%%%%%%%%%%%%%%
 I^{++}=&\eta c,\qquad
 I^{--}\ =\ b\xi,\qquad
 I\ =\ cb-\xi\eta.\label{topn4}
\end{align} 
The conditions (\ref{pscorg}) can be written 
in terms of these $N=4$ generators as
\begin{subequations}\label{psc1}
\begin{align}
 G^+_0|\psi\rangle=&0,\qquad 
\delta|\psi\rangle=G^+_0|\Lambda\rangle,\\
%%%%%%%%%%%%%%%%%%%%%%%%
I_0|\psi\rangle=&|\psi\rangle,\\
%%%%%%%%%%%%%%%%%%%%%%%%
\widetilde G^+_0|\psi\rangle=&\widetilde G^+_0|\Lambda\rangle=0,
\label{eta}
\end{align}
\end{subequations}
which are the definitions of the physical states
in the $N=4$ topological string theory.\cite{BV}
Because the $\eta_0$-cohomology is trivial, we can always
solve Eq.~(\ref{eta}) as $|\psi\rangle=\widetilde G^+_0|V\rangle$ and
$|\Lambda\rangle=\widetilde G^+_0|\Lambda^-\rangle$,
and this allows us 
to rewrite (\ref{psc1}) in the more symmetric forms
\begin{subequations}\label{phys}
\begin{align}
 G^+_0\widetilde G^+_0|V\rangle=&0,\label{eom}\\
%%%%%%%%%%%%%%%%%%%%%%%%%%%%%%%%%
\delta|V\rangle=&G^+_0|\Lambda^-\rangle
+ \widetilde G^+_0|\widetilde\Lambda^-\rangle,\label{gauge}\\
%%%%%%%%%%%%%%%%%%%%%%%%%%%%%%%%%
I_0|V\rangle=&0.\label{u1}
\end{align}
\end{subequations}
In this paper, we regard the first condition, (\ref{eom}), as
the equation of motion and the second, (\ref{gauge}), as
the gauge transformation, according to the standard terminology of 
string field theory. These conditions will be solved in order 
to investigate some physical states in \S\ref{physspec}.

\section{Hybrid superstrings on $AdS_3\times S^1\times\mathcal{N}/U(1)$}
\label{hybrid}

We develop the hybrid formalism on $AdS_3\times S^1$ in this section.
We introduce the hybrid fields through a field redefinition from
the RNS fields, which allows the entire space-time supersymmetry to be
manifest. Using these new fields, the space-time $N=2$ superconformal 
generators are constructed explicitly. 

The basic fields of the RNS superstrings on $AdS_3\times S^1$ are the
matter fields $(X^0,X^1,\phi_L,Y,\psi^0,$ $\psi^1,\psi^3,\psi^4)$ 
and the superconformal ghosts $(b,c,\beta,\gamma)$, as explained 
in the previous section. Here we use the bosonized forms 
$(H_0,H_1,H'_2,\phi,\chi,\sigma)$ for the world-sheet fermions
(\ref{psifb}), the $U(1)$ current (\ref{u1fb}) and 
the superconformal ghosts (\ref{bg}).
In order to introduce the hybrid fields, we first carry out 
a linear transformation on these six bosons as
\begin{align}
  \phi_{-+}=&-\frac{i}{2}H_0-\frac{i}{2}H_1
+\frac{i}{2}\sqrt{3}H'_2-\frac{3}{2}\phi+\chi+\sigma,\nonumber\\
%%%%%%%%%%%%%%%%%%%%%%%%%%%%
 \phi_{+-}=&\frac{i}{2}H_0-\frac{i}{2}H_1
-\frac{i}{2}\sqrt{3}H'_2+\frac{1}{2}\phi,\nonumber\\
%%%%%%%%%%%%%%%%%%%%%%%%%%%%
 \phi_{++}=&\frac{i}{2}H_0+\frac{i}{2}H_1
+\frac{i}{2}\sqrt{3}H'_2-\frac{3}{2}\phi+\chi+\sigma,\nonumber\\
%%%%%%%%%%%%%%%%%%%%%%%%%%%%
 \phi_{--}=&-\frac{i}{2}H_0+\frac{i}{2}H_1
-\frac{i}{2}\sqrt{3}H'_2+\frac{1}{2}\phi,\nonumber\\
%%%%%%%%%%%%%%%%%%%%%%%%%%%%
i\rho=&\sqrt{3}iH'_2-3\phi+2\chi+\sigma,\nonumber\\
%%%%%%%%%%%%%%%%%%%%%%%%%%%%
\sqrt{3}i\widehat H'_2=&\sqrt{3}iH'_2
-3\phi+3\chi.
\end{align}
These relations yield
\begin{align}
 i\widehat H_2=&iH_2-\alpha(\phi-\chi),\nonumber\\
 i\widehat Y=&iY-Q(\phi-\chi).
\end{align}
Then, we define the space-time spinor fields and their conjugates as
\begin{alignat}{2}
 \Theta^\alpha=&k^{-\frac{1}{4}}e^{\phi_{\alpha+}},&\qquad
 \bar\Theta^\alpha=&k^{-\frac{1}{4}}e^{\phi_{\alpha-}},\nonumber\\
%%%%%%%%%%%%%%%%%%%
 \mathcal{P}_\alpha=&k^{\frac{1}{4}}e^{-\phi_{\alpha+}},&\qquad
 \bar{\mathcal{P}}_\alpha=&k^{\frac{1}{4}}e^{-\phi_{\alpha-}},\qquad
(\alpha=\pm)
\end{alignat}
satisfying
\begin{equation}
 \Theta^\alpha(z)\mathcal{P}_\beta(w)\sim
\frac{\delta^\alpha_\beta}{z-w},\qquad
 \bar\Theta^\alpha(z)\bar{\mathcal{P}}_\beta(w)\sim
\frac{\delta^\alpha_\beta}{z-w}.
\end{equation}
The basic fields in the hybrid formalism are finally obtained as
the superspace coordinates (and conjugates of fermionic coordinates)
 with an additional boson:
$(X^0,X^1,\phi_L,\widehat Y;\Theta^\alpha,\bar\Theta^\alpha,
\mathcal{P}_\alpha,\bar{\mathcal{P}}_\alpha,\rho)$. 
The $U(1)$ boson in the $\mathcal{N}/U(1)$ sector is modified to
$\widehat H_2$, which requires modifications of the $N=2$ superconformal 
generators to 
$(\widehat T_{\mathcal{N}/U(1)},\widehat G^{\pm}_{\mathcal{N}/U(1)},
\widehat I_{\mathcal{N}/U(1)})$, which are uniquely determined through
the relation
\begin{equation}
 \widehat I_{\mathcal{N}/U(1)}=-\alpha i\partial\widehat H_2.
\end{equation}
We note here that these new generators of the $\mathcal{N}/U(1)$ sector
completely (anti-) commute with the hybrid fields in the $AdS_3\times
S^1$ sector.

In terms of these hybrid fields, 
the space-time supercharges are written
\begin{align}
  \mathcal{G}^+_{\pm\frac{1}{2}}=&\oint\dz\bar{\mathcal{P}}_\mp,\nonumber\\
%%%%%%%%%%%%%%%%%
 \mathcal{G}^-_{\pm\frac{1}{2}}=&\oint\dz\Bigg(
\mathcal{P}_\mp+\bar\Theta^\mp e^{\mp\beta i( X^0+ X^1)}
(\frac{1}{\beta}i\partial X^1\mp\frac{1}{Q}\partial\phi_L)\nonumber\\
&\hspace{15mm}
-\bar\Theta^\pm(\frac{1}{\beta}i\partial X^0\pm
\frac{1}{Q}i\partial\widehat Y\mp i\partial\rho
\pm\Theta^\pm\mathcal{P}_\pm\mp\bar\Theta^\mp\bar{\mathcal{P}}_\mp)
%\nonumber\\
%&\hspace{15mm}
\pm\Theta^\pm\bar\Theta^\mp\mathcal{P}_\mp
\Bigg),
%%%%%%%%%%%%%%%%%
\end{align}
where we have carried out the similarity transformation generated by
\begin{equation}
 R_0=-\oint\dz\left(
\frac{\sqrt{2}}{Q}e^{-i\rho}\Theta^+\Theta^-\widehat G^-_{\mathcal{N}/U(1)}
\right),
\end{equation}
so that the supersymmetry may be closed in the $AdS_3\times S^1$ sector.
In our case, this space-time supersymmetry is enlarged to the $N=2$
superconformal symmetry with $c=6kp$. We can explicitly construct
their generators as
\begin{subequations}\label{scharge}
\begin{align}
 \mathcal{L}_n=&\oint\dz\Bigg(
\gamma^n\left(
-\frac{1}{\beta}i\partial X^0-n\frac{1}{Q}\partial\phi_L\right)
%\nonumber\\
%&\hspace{10mm} %%%%%%%%%%%%
+\frac{1}{2}(n^2-1)\gamma^n
\left(\Theta^+\mathcal{P}_+-\Theta^-\mathcal{P}_-+\bar\Theta^+\bar{\mathcal{P}}_+
-\bar\Theta^-\bar{\mathcal{P}}_-\right)
\nonumber\\
&\hspace{10mm} %%%%%%%%%%%%
-\frac{1}{2}n(n+1)\gamma^{n-1}
\left(\Theta^+\mathcal{P}_-+\bar\Theta^+\bar{\mathcal{P}}_-\right)
%\nonumber\\
%&&\hspace{10mm}
+\frac{1}{2}n(n-1)\gamma^{n+1}
(\Theta^-\mathcal{P}_++\bar\Theta^-\bar{\mathcal{P}}_+)
\Bigg),\label{stvir}\\
%%%%%%%%%%%%%%%%%%%
\mathcal{G}^+_r=&\oint\dz\Bigg(
\left(\frac{1}{2}-r\right)\gamma^{r+\frac{1}{2}}\bar{\mathcal{P}}_+
+\left(\frac{1}{2}+r\right)\gamma^{r-\frac{1}{2}}\bar{\mathcal{P}}_-
\Bigg),\label{stgp}\\
%%%%%%%%%%%%%%%%%%%
\mathcal{G}^-_r=&\oint\dz\Bigg(
\left(\frac{1}{2}-r\right)\gamma^{r+\frac{1}{2}}\bigg(
\mathcal{P}_++\bar\Theta^+e^{\beta i( X^0+ X^1)}
\left(\frac{1}{\beta}i\partial X^1+\frac{1}{Q}\partial\phi_L\right)
\nonumber\\
&\hspace{10mm}
-\bar\Theta^-\left(\frac{1}{\beta}i\partial X^0
-\frac{1}{Q}i\partial\widehat Y+i\partial\rho
+\left(r+\frac{1}{2}\right)\Theta^+\mathcal{P}_+
-\left(r+\frac{3}{2}\right)\Theta^-\mathcal{P}_-+\bar\Theta^+\bar{\mathcal{P}}_+
\right)
\nonumber\\
&\hspace{10mm}
+\left(r-\frac{1}{2}\right)\Theta^-\bar\Theta^+\mathcal{P}_+-\partial\bar\Theta^-
\bigg)\nonumber\\
&\hspace{10mm}
+\left(\frac{1}{2}+r\right)\gamma^{r-\frac{1}{2}}\bigg(
\mathcal{P}_-+\bar\Theta^-e^{-\beta i( X^0+ X^1)}
\left(\frac{1}{\beta}i\partial X^1-\frac{1}{Q}\partial\phi_L\right)\nonumber\\
&\hspace{10mm}
-\bar\Theta^+\left(\frac{1}{\beta}i\partial X^0+\frac{1}{Q}i\partial\widehat Y
-i\partial\rho
-\left(r-\frac{3}{2}\right)\Theta^+\mathcal{P}_+
+\left(r-\frac{1}{2}\right)\Theta^-\mathcal{P}_-
-\bar\Theta^-\bar{\mathcal{P}}_-\right)\nonumber\\
&\hspace{10mm}
+\left(r+\frac{1}{2}\right)\Theta^+\bar\Theta^-\mathcal{P}_-+\partial\bar\Theta^+
\bigg)
\nonumber\\
&\hspace{10mm}
-\left(r^2-\frac{1}{4}\right)\left(
\gamma^{r+\frac{3}{2}}\Theta^-\bar\Theta^-\mathcal{P}_+
-\gamma^{r-\frac{3}{2}}\Theta^+\bar\Theta^+\mathcal{P}_-\right)\Bigg),
\label{stgm}\\
%%%%%%%%%%%%%%%%%%%
\mathcal{I}_n=&\oint\dz\Bigg(
\gamma^n
\left(2\left(\frac{1}{Q}i\partial\widehat Y-i\partial\rho\right)
+\left(\Theta^+\mathcal{P}_++\Theta^-\mathcal{P}_--\bar\Theta^+\bar{\mathcal{P}}_+
-\bar\Theta^-\bar{\mathcal{P}}_-\right)\right)
\nonumber\\
&\hspace{10mm}
-n\gamma^n\left(\Theta^+\mathcal{P}_+-\Theta^-\mathcal{P}_-
-\bar\Theta^+\bar{\mathcal{P}}_+
+\bar\Theta^-\bar{\mathcal{P}}_-\right)
\nonumber\\
&\hspace{10mm}
+n\gamma^{n-1}\left(\Theta^+\mathcal{P}_--\bar\Theta^+\bar{\mathcal{P}}_-\right)
-n\gamma^{n+1}\left(\Theta^-\mathcal{P}_+-\bar\Theta^-\bar{\mathcal{P}}_+\right)
\Bigg),\label{stu1}
\end{align}
\end{subequations}
where $\gamma=e^{-\beta i(X^0+X^1)}$.
We use these explicit forms to obtain the transformation laws of some 
lower-level physical fields in \S\ref{stsusy}.

In order to obtain the physical state conditions,
we must also rewrite the $N=4$ topological superconformal generators
(\ref{topn4}) in terms of the hybrid fields. Here we present only the $N=2$ 
subset, in the forms they take after the similarity transformation 
$R_0$, which are necessary and sufficient to obtain all the $N=4$ generators:
\begin{align}
  T=&-\frac{1}{2}\partial X^\mu\partial X_\mu
-\frac{1}{2}\partial\phi_L\partial\phi_L-\frac{Q}{2}\partial^2\phi_L
-\frac{1}{2}\partial\widehat Y\partial\widehat Y
-\frac{Q}{2}i\partial^2\widehat Y
\nonumber\\
&
-\mathcal{P}_\alpha\partial\Theta^\alpha
-\bar{\mathcal{P}}_\alpha\partial\bar\Theta^\alpha
+\frac{1}{2}\partial\rho\partial\rho
+\frac{1}{2}i\partial^2\rho
+\widehat T_{\mathcal{N}/U(1)}+\frac{1}{2}\partial\widehat
 I_{\mathcal{N}/U(1)},
\nonumber\\
%%%%%%%%%%%%%%%
G^+=&\frac{Q}{\sqrt{2}}{}^\times_\times e^{-i\rho}\left(
1+\Theta^+\bar\Theta^--\Theta^-\bar\Theta^+\right)\bar d_+
\bar d_-{}^\times_\times +\widehat G^+_{\mathcal{N}/U(1)},
\nonumber\\
%%%%%%%%%%%%%%%
G^-=&\frac{Q}{\sqrt{2}}e^{i\rho}d_+
d_-+\widehat G^-_{\mathcal{N}/U(1)},\nonumber\\
%%%%%%%%%%%%%%%
I=&i\partial\rho-Qi\partial\widehat Y+\widehat I_{\mathcal{N}/U(1)},
\end{align}
where
\begin{align}
 d_\mp=&\mathcal{P}_\mp,\nonumber\\
%%%%%%%%%%%%%%%%%%%%%%%%%%%%%%%%%%%%%%
 \bar d_\pm=&\bar{\mathcal{P}}_\pm-\Theta^\pm e^{\pm\beta i( X^0+ X^1)}
\left(\frac{1}{\beta}i\partial X^1\pm\frac{1}{Q}\partial\phi_L\right)
\nonumber\\
&
+\Theta^\mp\left(\frac{1}{\beta}i\partial X^0
\pm\frac{1}{Q}i\partial\widehat Y\mp i\partial\rho
\pm\Theta^\pm\mathcal{P}_\pm\mp\bar\Theta^\mp\bar{\mathcal{P}}_\mp\right)
\nonumber\\
&
\mp\Theta^\pm\bar\Theta^\mp\bar{\mathcal{P}}_\pm
+2\Theta^\pm\Theta^\mp\bar\Theta^\mp\left(
\frac{1}{Q}i\partial\widehat Y-i\partial\rho\right)
\nonumber\\
&
+\frac{2}{Q^2}\Theta^\pm\Theta^\mp\partial\bar\Theta^\mp
+2\partial\left(\Theta^\pm\Theta^\mp\right)\bar\Theta^\mp,
\end{align}
and $\norm{\ }$ denotes normal ordering with respect to
the coefficient fields and the {\it currents} $d_\alpha$
and $\bar d_\alpha$.
The zero modes $G^+_0,\ \widetilde G^+_0$ and $I_0$ are
(anti-)commutative with the supercharges (\ref{scharge}).
This guarantees a supersymmetric physical spectrum.
Although these forms are useful for explicit calculation,
and will be used in \S\ref{physspec}, the hybrid fields possess
some non-trivial hermiticity properties.
We must carry out a further similarity transformation 
to obtain fields with conventional hermiticity.
We present this similarity transformation in Appendix B.

\section{Spectral flow and the Hilbert space of the hybrid superstring}
\label{flow}

Now, let us study the Hilbert space of the hybrid superstring, including
spectrally flowed representations.\cite{MO} 
We consider the structure of the $sl(2)$ currents (\ref{slcurrent}),
because the spectral flow is defined by their automorphism.
They can be decomposed into three independent parts as
\begin{equation}
 J^a=j^a+\sum_{I=1}^2J_{(I)}^a,\label{totalsl2}
\end{equation}
where $j^a$ represents the bosonic currents given in (\ref{bsl2}), and
 \begin{alignat}{2}
  J_{(1)}^{\pm\pm}=&\pm\Theta^\pm\mathcal{P}_\mp,\quad & 
  J_{(1)}^3=&\frac{1}{2}(\Theta^+\mathcal{P}_+-\Theta^-\mathcal{P}_-),\nonumber\\
%%%%%%%%%%%%%%%%%%%%%%%%%%%%%%%%%%%%%%
  J_{(2)}^{\pm\pm}=&\pm\bar\Theta^\pm\bar{\mathcal{P}}_\mp, \quad &
  J_{(2)}^3=&\frac{1}{2}(\bar\Theta^+\bar{\mathcal{P}}_+-\bar\Theta^-\bar{\mathcal{P}}_-). 
 \end{alignat}
are the two independent fermionic currents of level $-1$.

The flowed representation of the bosonic part is defined by\cite{MO}
 \begin{align}
j^3_0|j,m,p\rangle_0=&\left(m-\left(\frac{k+2}{2}\right)p\right)
|j,m,p\rangle_0,\nonumber\\
%%%%%%%%%%%
j^{\pm\pm}_{\mp p}|j,m,p\rangle_0=&(m\pm
  j)|j,m\pm1,p\rangle_0,\nonumber\\
%%%%%%%%%%%
j^3_n|j,m,p\rangle_0=&0,\qquad
j^{\pm\pm}_{n\mp p}|j,m,p\rangle_0=0,\qquad (n>0)\label{sl2cond}
 \end{align}
which is realized as the oscillator vacuum of the free-field realization
of the discrete light-cone Liouville theory as\cite{HHS}
\begin{equation}
 |j,m,p\rangle_0=e^{-i(m\beta-\frac{p}{\beta})X^0
-im\beta X^1-Qj\phi_L}(0)|0\rangle_B,
\end{equation}
where $|0\rangle_B$ is the bosonic $SL(2,R)$ invariant vacuum.
For string theory on $AdS_3$, we must include
all the spectrally flowed representations of the continuous 
$\hat{\cal C}^{\alpha(p)}_{\frac{1}{2}+is}$ and
the discrete representation $\hat{\cal D}^{+(p)}_{j}$ 
for consistency.\cite{MO,HHS}

Now, the extension to the fermionic part is obtained by simply 
replacing $j^a$ with the fermionic current $J^a_{(1)}$ or $J^a_{(2)}$ 
and assuming that the fermionic $SL(2,R)$ invariant vacuum is 
its singlet. These conditions are realized by setting $j=m=0$ and 
replacing $k+2\rightarrow-1$ in (\ref{sl2cond}). 
The flowed representation is then given by the state
$|p\rangle_F$, defined by
 \begin{alignat}{4}
\Theta^+_{n-p}|p\rangle_F=&0, &\qquad
(\mathcal{P}_+)_{n+p}|p\rangle_F=&0,&\qquad &(n>0)\nonumber\\
%%%%%%%%%%%%%%%%%%%%%%%%%%
\Theta^-_{n}|p\rangle_F=&0, &\qquad
(\mathcal{P}_-)_{n-1}|p\rangle_F=&0,&\qquad &(n>0)\nonumber\\
%%%%%%%%%%%%%%%%%%%%%%%%%%
\bar\Theta^+_{n}|p\rangle_F=&0,&\qquad
(\bar{\mathcal{P}}_+)_{n-1}|p\rangle_F=&0,&\qquad &(n>0)\nonumber\\
%%%%%%%%%%%%%%%%%%%%%%%%%%
\bar\Theta^-_{n+p}|p\rangle_F=&0,&\qquad
(\bar{\mathcal{P}}_-)_{n-1-p}|p\rangle_F=&0,&\qquad &(n>0)
 \end{alignat}
which is explicitly constructed (up to a sign) 
on the fermionic $SL(2,R)$ invariant vacuum 
$|p=0\rangle_F$ as
\begin{equation}
|p\rangle_F=\left\{
 \begin{array}{ll}
\prod_{r=0}^{p-1}\Theta^+_{-r}\prod_{s=1}^{p}(\bar{\mathcal{P}}_-)_{-s}
|0\rangle_F
& \textrm{for $p>0$},\\
%%%%%%%%%%%%%%%%%%%%%%%%%%
\prod_{r=1}^{-p}(\mathcal{P}_+)_{-r}\prod_{s=0}^{-p-1}\bar\Theta^-_{-s}
|0\rangle_F
& \textrm{for $p<0$}.
 \end{array}
\right.\label{fock}
\end{equation}
Moreover, we use the representation 
 \begin{align}
 \Theta^+_{-p}|\boldsymbol{\theta},p\rangle_F
=&|\boldsymbol{\theta},p\rangle_F\theta^+,&\qquad
(\mathcal{P}_+)_p|\boldsymbol{\theta},p\rangle_F
=&|\boldsymbol{\theta},p\rangle_F\frac{\partial}{\partial\theta^+},\nonumber\\
%%%%%%%%%%%%%%%%%%%%%%%%%%
 \Theta^-_0|\boldsymbol{\theta},p\rangle_F
=&|\boldsymbol{\theta},p\rangle_F\theta^-,&\qquad
(\mathcal{P}_-)_0|\boldsymbol{\theta},p\rangle_F
=&|\boldsymbol{\theta},p\rangle_F\frac{\partial}{\partial\theta^-},\nonumber\\
%%%%%%%%%%%%%%%%%%%%%%%%%%
 \bar\Theta^+_0|\boldsymbol{\theta},p\rangle_F
=&|\boldsymbol{\theta},p\rangle_F\bar\theta^+,&\qquad
(\bar{\mathcal{P}}_+)_0|\boldsymbol{\theta},p\rangle_F
=&|\boldsymbol{\theta},p\rangle_F\frac{\partial}
{\partial\bar\theta^+},\nonumber\\
%%%%%%%%%%%%%%%%%%%%%%%%%%
 \bar\Theta^-_p|\boldsymbol{\theta},p\rangle_F
=&|\boldsymbol{\theta},p\rangle_F\bar\theta^-,&\qquad
(\bar{\mathcal{P}}_-)_{-p}|\boldsymbol{\theta},p\rangle_F
=&|\boldsymbol{\theta},p\rangle_F\frac{\partial}{\partial\bar\theta^-}
 \end{align}
for the fermionic \lq\lq zero-modes'',
where $\boldsymbol{\theta}=(\theta^\pm,\bar{\theta}^\pm)$ are
the fermionic {\it coordinates}. An explicit expression of this state
$|\boldsymbol{\theta},p\rangle_F$ is obtained from the oscillator
state (\ref{fock}) as
\begin{equation}
 |\boldsymbol{\theta},p\rangle_F
=(\Theta^+_{-p}-\theta^+)(\Theta^-_0-\theta^-)
(\bar{\Theta}^+_0-\bar{\theta}^+)(\bar{\Theta}^-_p-\bar{\theta}^-)|p\rangle_F.
\end{equation}
Eventually, the spectrally flowed representation of 
the total algebra (\ref{totalsl2}) is obtained as the tensor
product $|j,m,p,\boldsymbol{\theta}\rangle
=|j,m,p\rangle_0\otimes|\boldsymbol{\theta},p\rangle_F$.

For the $S^1$ direction $\widehat Y$ and the additional boson $\rho$, 
we take the oscillator ground states as
\begin{equation}
 |q,l\rangle=e^{iQq\widehat Y-i(l+p)\rho}(0)|0\rangle,
\end{equation}
where the eigenvalue $l+p$ of the $\rho$-momentum 
is determined such that the physical state conditions
(\ref{phys}) are consistent with the spectral flow.
The unflowed part $l$ must be restricted to $l=0,\pm 1$ 
to avoid the infinite degeneracy due to the picture ambiguity 
in the RNS formalism.

Together with the representations $|\Delta_\mathcal{N},Q_\mathcal{N}\rangle$
of the $N=2$ superconformal algebra in the $\mathcal{N}/U(1)$ sector 
characterized by
 \begin{align}
L_0^{\mathcal{N}/U(1)}|\Delta_\mathcal{N},Q_\mathcal{N}\rangle
=&\Delta_\mathcal{N}|\Delta_\mathcal{N},Q_\mathcal{N}\rangle,\nonumber\\
%%%%%%%%%%%%%%%%%%%%%%%%%
J_0^{\mathcal{N}/U(1)}|\Delta_\mathcal{N},Q_\mathcal{N}\rangle
=&Q_\mathcal{N}|\Delta_\mathcal{N},Q_\mathcal{N}\rangle,
 \end{align}
we can define the total Hilbert space of the hybrid superstrings on
$AdS_3\times S^1\times\mathcal{N}/U(1)$ as the tensor product 
of all these ground states,
\begin{equation}
 |\boldsymbol{j},\boldsymbol{\theta},l,\Delta_\mathcal{N},Q_\mathcal{N}\rangle
=|j,m,p\rangle_0\otimes|\boldsymbol{\theta},p\rangle_F 
\otimes|q,l\rangle\otimes|\Delta_\mathcal{N},Q_\mathcal{N}\rangle,
\end{equation}
where the zero-modes $(j,m,p,q)$ are denoted simply by $\boldsymbol{j}$. 
It is useful to note that the world-sheet energy of this state is given by
\begin{equation}
 L_0|\boldsymbol{j},\boldsymbol{\theta},l,\Delta_\mathcal{N},Q_\mathcal{N}\rangle=
\left(-\frac{j(j-1)}{k}+(m-l)p-\frac{k}{4}p^2
+\frac{1}{k}q^2+\Delta_\mathcal{N}-\frac{l^2}{2}\right)
|\boldsymbol{j},\boldsymbol{\theta},l,\Delta_\mathcal{N},Q_\mathcal{N}\rangle,
\end{equation}
where we use the $U(1)$ condition (\ref{u1}), which now leads to
$l+p-Q^2q+Q_\mathcal{N}=0$.

\section{Physical spectrum}\label{physspec}

In this section, we explicitly investigate the physical spectrum 
for the oscillator ground states constructed in the previous section. 
We concentrate on the states whose $\mathcal{N}/U(1)$ 
sector is the world-sheet chiral primary states characterized by 
$\Delta_\mathcal{N}=\frac{Q_\mathcal{N}}{2}$. In this case,
we find physical states for $l=-1$ ($l=0$) 
identified with the first (second) series
of the space-time chiral primaries,
which have been obtained in the RNS formalism.\cite{HHS}$^,$
\footnote{
We conjecture that there is no physical state for $l=1$ in this sector. 
However, this is not explicitly shown in this paper. }

Let us first consider the $l=-1$ case in detail.
The oscillator ground state is generally given by
\begin{equation}
|V\rangle=|\boldsymbol{j},\boldsymbol{\theta},-1\rangle
\Phi(\boldsymbol{j},\boldsymbol{\theta}), 
\end{equation}
where $\Phi$ is the superfield, which is a function
of zero-modes $(\boldsymbol{j},\boldsymbol{\theta})
=(j,m,p,q,\theta^\pm,\bar\theta^\pm)$. The eigenvalues of the
$\mathcal{N}/U(1)$ sector are omitted, because they are not independent,
due to the U(1) condition $Q_\mathcal{N}=1-p+Q^2q$
and the chirality $\Delta_\mathcal{N}=\frac{Q_\mathcal{N}}{2}$. 
In terms of this superfield $\Phi$, the equation of motion (\ref{eom}) 
and the gauge transformation (\ref{gauge}) can be written
\begin{subequations}
 \begin{align}
   D_-D_+{\cal T}_0\bar D_+\bar D_-\Phi=&0,\label{eom-1}\\
   \delta\Phi=&\bar D_+\bar\Lambda^++\bar D_-\bar\Lambda^-,\label{gauge-1}
 \end{align}
\end{subequations}
where
 \begin{align}
   {\cal T}_0=&1+\theta^+\bar\theta^--\theta^-\bar\theta^+,\nonumber\\
%%%%%%%%%%%%%%%%%%%%%
D_+=&\frac{\partial}{\partial\theta^+},\qquad
%%%%%
D_-=\frac{\partial}{\partial\theta^-},\nonumber\\
%%%%%%%%%%%%%%%%%%%%%
\bar D_+=&\frac{\partial}{\partial{\bar\theta}^+}
+\theta^+\nabla^{--}+\theta^-
\left(\nabla^3+q-l\right)+\theta^-\theta^+
\frac{\partial}{\partial\theta^+}
\nonumber\\
&\hspace{10mm}
-\theta^-\bar\theta^-
\frac{\partial}{\partial{\bar\theta}^-}
-\theta^+\bar\theta^-\frac{\partial}{\partial{\bar\theta}^+}
-2\theta^-\theta^+\bar\theta^-
(q-l-\frac{k}{2}p),\nonumber\\
%%%%%%%%%%%%%%%%%%%%%
\bar D_-=&\frac{\partial}{\partial{\bar\theta}^-}
+\theta^-\nabla^{++}+\theta^+\left(
\nabla^3-q+l\right)-\theta^+\theta^-
\frac{\partial}{\partial\theta^-}
\nonumber\\
&\hspace{10mm}
+\theta^+\bar\theta^+
\frac{\partial}{\partial{\bar\theta}^+}
+\theta^-\bar\theta^+
\frac{\partial}{\partial{\bar\theta}^-}
-2\theta^-\bar\theta^+\theta^+(q-l).
 \end{align}
The operators $(\nabla^{\pm\pm},\nabla^3)$ act on
$\Phi(\boldsymbol{j},\boldsymbol{\theta})$ as
 \begin{align}
\nabla^{\pm\pm}\Phi(j,m,p,q,\boldsymbol{\theta})=&
(m\mp1\pm j)\Phi(j,m\mp1,p,q,\boldsymbol{\theta}),\nonumber\\
%%%%%%%%%%%%%%%
\nabla^3\Phi(\boldsymbol{j},\boldsymbol{\theta})=&
\left(m-\frac{k}{2}p\right)\Phi(\boldsymbol{j},\boldsymbol{\theta}).
 \end{align}
We can obtain the gauge transformation (\ref{gauge-1})
by taking the gauge parameter state as, e.g.,
\begin{equation}
|\Lambda^-\rangle=
\bar\Theta^-_{-1+p}|\boldsymbol{j},\boldsymbol{\theta},-2\rangle
{\bar\Lambda}^+(\boldsymbol{j},\boldsymbol{\theta})+
\bar\Theta^+_{-1}|\boldsymbol{j},\boldsymbol{\theta},-2\rangle
{\bar\Lambda}^-(\boldsymbol{j},\boldsymbol{\theta}),
\end{equation}
where the quantities ${\bar\Lambda}^\pm(\boldsymbol{j},\boldsymbol{\theta})$
are the gauge parameter superfields.

In order to solve the equation of motion (\ref{eom-1})
and find an explicit form of the physical spectrum,
we expand the superfields $\Phi$ and $\bar{\Lambda^\pm}$ as
\begin{subequations}
 \begin{align}
   \Phi=&\phi
+\theta^+\bar\psi_+
+\theta^-\bar\psi_-
+\bar\theta^+\psi_+
+\bar\theta^-\psi_-
\nonumber\\
 &      %%%%%%%%%%%%%%%%%%%%%%%
+\theta^+\bar\theta^+v_{++}
+\theta^-\bar\theta^+(v_3+v_Y)
+\theta^+\bar\theta^-(v_3-v_Y)
+\theta^-\bar\theta^-v_{--}
+\theta^+\theta^-\omega
+\bar\theta^+\bar\theta^-\varphi
\nonumber\\
&      %%%%%%%%%%%%%%%%%%%%%%%
+\theta^-\bar\theta^+\bar\theta^-\chi^+
-\theta^+\bar\theta^+\bar\theta^-\chi^-
-\theta^+\theta^-\bar\theta^-\bar\chi^+
+\theta^+\theta^-\bar\theta^+\bar\chi^-
+\theta^+\theta^-\bar\theta^+\bar\theta^-d,\label{pcomp}\\
%%%%%%%%%%%%%%%%%%%%%%%%%%%%%%
 \bar\Lambda^+=&
\bar\epsilon^+
+\theta^+\alpha^+_+
+\theta^-\alpha^+_-
+\bar\theta^+\lambda^+_+
+\bar\theta^-\lambda^+_-
\nonumber\\
&      %%%%%%%%%%%%%%%%%%%%%%%
-\theta^+\bar\theta^+\bar\eta^+_{++}
+\theta^-\bar\theta^+\bar\eta^+_{-+}
+\theta^+\bar\theta^-\bar\eta^+_{+-}
+\theta^-\bar\theta^-\bar\eta^+_{--}
-\theta^+\theta^-\zeta^+_{-+}
+\bar\theta^+\bar\theta^-\eta^+_{+-}
\nonumber\\
&      %%%%%%%%%%%%%%%%%%%%%%%
+\theta^-\bar\theta^+\bar\theta^-f^{++}
-\theta^+\bar\theta^+\bar\theta^-f^{+-}
-\theta^+\theta^-\bar\theta^-m^{++}
+\theta^+\theta^-\bar\theta^+m^{+-}
+\theta^+\theta^-\bar\theta^+\bar\theta^-\bar\xi^+,\\
%%%%%%%%%%%%%%%%%%%%%%%%%%%%%%%
 \bar\Lambda^-=&
\bar\epsilon^-
+\theta^+\alpha^-_+
+\theta^-\alpha^-_-
+\bar\theta^+\lambda^-_+
+\bar\theta^-\lambda^-_-
\nonumber\\
&      %%%%%%%%%%%%%%%%%%%%%%%
-\theta^+\bar\theta^+\bar\eta^-_{++}
+\theta^-\bar\theta^+\bar\eta^-_{-+}
+\theta^+\bar\theta^-\bar\eta^-_{+-}
+\theta^-\bar\theta^-\bar\eta^-_{--}
-\theta^+\theta^-\zeta^-_{-+}
+\bar\theta^+\bar\theta^-\eta^-_{+-}
\nonumber\\
&      %%%%%%%%%%%%%%%%%%%%%%%
+\theta^-\bar\theta^+\bar\theta^- f^{-+}
-\theta^+\bar\theta^+\bar\theta^- f^{--}
-\theta^+\theta^-\bar\theta^-m^{-+}
+\theta^+\theta^-\bar\theta^+m^{--}
+\theta^+\theta^-\bar\theta^+\bar\theta^-\bar\xi^-,
 \end{align}
\end{subequations}
where the component fields are functions of 
$\boldsymbol{j}=(j,m,p,q)$. 
The gauge transformation (\ref{gauge-1}) can be given
in terms of these component fields as
 \begin{align}
 \delta\phi=&\lambda^+_++\lambda^-_-,\nonumber\\
%%%%%%%%%%%%%%
\delta\psi_+=&-\eta^-_{+-},\nonumber\\
%%%%%%%%%%%%%
\delta\psi_-=&\eta^+_{+-},\nonumber\\
%%%%%%%%%%%%%
\delta\bar\psi_+=&\bar\eta^+_{++}-\bar\eta^-_{+-}
+\nabla^{--}\bar\epsilon^++(\nabla^3-q-1)\bar\epsilon^-,
\nonumber\\
%%%%%%%%%%%%%
 \delta\bar\psi_-=&-\bar\eta^+_{-+}-\bar\eta^-_{--}
+(\nabla^3+q+1)\bar\epsilon^++\nabla^{++}\bar\epsilon^-,
\nonumber\\
%%%%%%%%%%%%%%
\delta v_{++}=&-f^{--}+\nabla^{--}\lambda^+_+
+(\nabla^3-q)\lambda^-_+,\nonumber\\
%%%%%%%%%%%%%
\delta v_3=&\frac{1}{2}(f^{-+}+f^{+-})
+\frac{1}{2}\nabla^{++}\lambda^-_+
+\frac{1}{2}\nabla^{--}\lambda^+_-
+\frac{1}{2}\nabla^3(\lambda^+_++\lambda^-_-)
+\frac{1}{2}q(\lambda^+_+-\lambda^-_-),\nonumber\\
%%%%%%%%%%%%%
\delta v_{--}=&-f^{++}+(\nabla^3+q)\lambda^+_-
+\nabla^{++}\lambda^-_-,\nonumber\\
%%%%%%%%%%%%%
\delta v_Y=&\frac{1}{2}(f^{-+}-f^{+-})
+\frac{1}{2}\nabla^{++}\lambda^-_+
-\frac{1}{2}\nabla^{--}\lambda^+_-\nonumber\\
&\hspace{20mm}
+\frac{1}{2}\nabla^3(\lambda^+_+-\lambda^-_-)
+\frac{1}{2}(q+2)(\lambda^+_++\lambda^-_-),\nonumber\\
%%%%%%%%%%%%
\delta\omega=&
m^{+-}-m^{-+}
+\nabla^{--}\alpha^+_{-}
-\nabla^{++}\alpha^-_+
-(\nabla^3+q+2)\alpha^+_+
+(\nabla^3-q-2)\alpha^-_-,\nonumber\\
%%%%%%%%%%%%
\delta\varphi=&0,\nonumber\\
%%%%%%%%%%%%
\delta\chi^+=&(\nabla^3+q)\eta^+_{+-}
+\nabla^{++}\eta^-_{+-},\nonumber\\
%%%%%%%%%%%%%%
\delta\chi^-=&-\nabla^{--}\eta^+_{+-}
-(\nabla^3-q)\eta^-_{+-},\nonumber\\
%%%%%%%%%%%%%
\delta\bar\chi^+=&-\bar\xi^+-\nabla^{--}\bar\eta^+_{--}
+(\nabla^3+q+1)\bar\eta^+_{+-}
\nonumber\\
&\hspace{20mm}
+\bar\eta^+_{-+}
-2(q+1-\frac{k}{2}p)\bar\epsilon^+
+\nabla^{++}\bar\eta^-_{+-}
-(\nabla^3-q-2)\bar\eta^-_{--},\nonumber\\
%%%%%%%%%%%%%%
\delta\bar\chi^-=&-\bar\xi^-+\nabla^{--}\bar\eta^+_{-+}
+(\nabla^3+q+2)\bar\eta^+_{++}
\nonumber\\
&\hspace{20mm}
+\nabla^{++}\bar\eta^-_{++}
+(\nabla^3-q-1)\bar\eta^-_{-+}-\bar\eta^-_{+-}
-2(q+1)\bar\epsilon^-,\nonumber\\
%%%%%%%%%%%%%%
\delta d=&\nabla^{--}f^{++}+\nabla^{++}f^{--}
+(\nabla^3+q+1)f^{+-}
\nonumber\\
&\hspace{10mm}
+(\nabla^3-q-1)f^{-+}
-2(q+1-\frac{k}{2}p)\lambda^+_+-2(q+1)\lambda^-_-.
 \end{align}
To fix these invariances, we choose the gauge conditions
 \begin{align}
& \phi=v_{++}=v_3=v_{--}=v_Y=\omega=0,\nonumber\\
&\psi_+=\psi_-=\bar\psi_+=\bar\psi_-=\bar\chi^+=\bar\chi^-=0.
\label{gcchiral}
 \end{align}
Then, the remaining fields form the {\it chiral} supermultiplet 
including two bosons and two fermions:
$\boldsymbol{\Phi}(m)=(\varphi(m),d(m-1);\chi^+(m),\chi^-(m-1))$.
Here and hereafter, we omit the $(j,p,q)$ dependence, because 
they are uniform in a given supermultiplet, that is, 
inert under supersymmetry transformations, as seen
in the next section.
The equations of motion for these reduced fields become
\begin{subequations}
 \begin{align}
&
d(m-1)=0,\\
&
\left({\cal K}_{(m)}+\Delta_\mathcal{N}-\frac{1}{2}\right)\varphi(m)=0,\label{KG}\\
&
\left(
\begin{array}{cc}
 m-\frac{k}{2}p-q-1&m-1+j \\
 m-j&m-\frac{k}{2}p+q \\
\end{array}
\right)
\left(
\begin{array}{c}
 \chi^+(m)\\
 \chi^-(m-1)\\
\end{array}\right)
=0,\label{Dirac}
 \end{align}
\end{subequations}
where ${\cal K}_{(m)}$ is the Klein-Gordon operator on 
$AdS_3\times S^1$, defined by
\begin{equation}
{\cal K}_{(m)}=
-\frac{j(j-1)}{k}+\frac{q^2}{k}+mp-\frac{k}{4}p^2, 
\end{equation}
where the last two terms, $mp$ and $-\frac{k}{4}p^2$,
are among the mass terms induced by the spectral flow. 
It should be noted that when $p\ne0$, this contribution 
depends on $m$, as indicated explicitly.
Equation~(\ref{Dirac}) is interpreted as the Dirac equation 
on $AdS_3\times S^1$.

Using the generators (\ref{stvir}) and (\ref{stu1}),
one can easily compute the space-time conformal weight and
the $R$-charge of the physical boson $\varphi$ as 
$\Delta_{s-t}=-m+\frac{k}{2}p$ and $Q_{s-t}=2q=k(Q_\mathcal{N}+p-1)$.
To compare this with the results in the RNS formalism,
we take $\mathcal{N}/U(1)=SU(2)/U(1)\times T^4$,
where the $SU(2)/U(1)$ sector is represented by the Kazama-Suzuki
model\cite{KS} with $c=3-\frac{6}{k}$. If we consider the ground 
state for the $T^4$ sector, the $U(1)$ charge spectrum of 
the world-sheet chiral primaries is given by 
$Q_\mathcal{N}=2\Delta_\mathcal{N}=s/k\ (s=0,1,\cdots,k-2)$.
Then the space-time $R$-charge becomes
\begin{equation}
 Q_{s-t}=s+(p-1)k,
\end{equation}
which coincides with the spectrum of the first series 
of the space-time chiral primaries obtained in Ref.~\citen{HHS}.
Actually, the space-time conformal weight can be independently 
calculated as
\begin{equation}
 \Delta_{s-t}=-m-\frac{k}{2}p=\frac{Q_{s-t}}{2}
\end{equation}
by taking $j=m=-q+\frac{k}{2}p=\frac{1}{2}(k-s)$
to solve the on-shell condition (\ref{KG}).\footnote{
The $p=0$ case is an exception. In that case, the on-shell condition
can be solved by setting $j=-q$, with $m$ arbitrary.
}
The fermionic partner can be obtained similarly by solving 
the Dirac equations given in (\ref{Dirac}). They form the on-shell chiral 
supermultiplet $(\varphi(m),\chi^+(m))$ of the space-time chiral primaries 
with $j=m=\frac{1}{2}(k-s)$.

For $l=0$, the $U(1)$ condition leads to
$q=\frac{1}{2}k(p+Q_\mathcal{N})$.  The equation
of motion (\ref{eom}) becomes
\begin{align}
 &
 \Bigg(
2D_-D_+{\cal T}_0\bar D_+\bar D_-
+D_+\bar D_-D_-{\cal T}_0\bar D_+
-D_-{\cal T}_0\bar D_-D_+\bar D_+\nonumber\\
&\hspace{10mm}
+D_-\bar D_+D_+{\cal T}_0\bar D_-
-D_+{\cal T}_0\bar D_+D_-\bar D_-
+D_+{\cal T}_0\bar D_-
-D_-{\cal T}_0\bar D_+
\Bigg)\Phi=0.\label{eoml0}
\end{align}
The explicit forms of the gauge transformation (\ref{gauge})
are different for the following two cases.
For $Q_\mathcal{N}\ne0$, it is given by
\begin{equation}
 \delta\Phi=\bar D_+\bar D_-{\cal T}_0\bar\Sigma,
\end{equation}
where the gauge parameter superfield comes from
\begin{equation}
 |\Lambda^-\rangle=|\boldsymbol{j},\boldsymbol{\theta},-1\rangle
\bar\Sigma(\boldsymbol{j},\boldsymbol{\theta}).
\end{equation}
The case $Q_\mathcal{N}=0$ corresponds to 
the ground state in the $\mathcal{N}/U(1)$ sector, 
which is often called \lq\lq compactification independent''. 
The gauge symmetry is enlarged in this case to
\begin{equation}
 \delta\Phi=\bar D_+\bar D_-{\cal T}_0\bar\Sigma
+D_+D_-\Sigma,\label{enl}
\end{equation}
where the additional transformation comes from the state
\begin{equation}
 |\Lambda^+\rangle=|j,m,p,q+1,\boldsymbol{\theta},2\rangle
\Sigma(\boldsymbol{j},\boldsymbol{\theta}),
\end{equation}
with $U(1)$ charge $-3+Q^2$ in the $\mathcal{N}/U(1)$ sector.

If we expand $\bar\Sigma$ and $\Sigma$ as
 \begin{align}
    \bar\Sigma=&
\bar\tau
+\theta^+\bar\zeta_+
+\theta^-\bar\zeta_-
+\bar\theta^+\bar\epsilon_+
+\bar\theta^-\bar\epsilon_-
\nonumber\\
&      %%%%%%%%%%%%%%%%%%%%%%%
-\theta^+\bar\theta^+\bar\lambda_{++}
+\theta^-\bar\theta^+\bar\lambda_{-+}
+\theta^+\bar\theta^-\bar\lambda_{+-}
+\theta^-\bar\theta^-\bar\lambda_{--}
-\theta^+\theta^-\bar\sigma
+\bar\theta^+\bar\theta^-\bar\lambda
\nonumber\\
&      %%%%%%%%%%%%%%%%%%%%%%%
+\theta^-\bar\theta^+\bar\theta^-\bar\eta^+
-\theta^+\bar\theta^+\bar\theta^-\bar\eta^-
-\theta^+\theta^-\bar\theta^-\bar\xi^+
+\theta^+\theta^-\bar\theta^+\bar\xi^-
+\theta^+\theta^-\bar\theta^+\bar\theta^-\bar f,\nonumber\\
%%%%%%%%%%%%%%%%%%%%%%%%%%%%%%%
 \Sigma=&
\tau
+\theta^+\epsilon_+
+\theta^-\epsilon_-
+\bar\theta^+\zeta_+
+\bar\theta^-\zeta_-
\nonumber\\
&      %%%%%%%%%%%%%%%%%%%%%%%
-\theta^+\bar\theta^+\lambda_{++}
+\theta^-\bar\theta^+\lambda_{-+}
+\theta^+\bar\theta^-\lambda_{+-}
+\theta^-\bar\theta^-\lambda_{--}
-\theta^+\theta^-\lambda
+\bar\theta^+\bar\theta^-\sigma
\nonumber\\
&      %%%%%%%%%%%%%%%%%%%%%%%
+\theta^-\bar\theta^+\bar\theta^-\xi^+
-\theta^+\bar\theta^+\bar\theta^-\xi^-
-\theta^+\theta^-\bar\theta^-\eta^+
+\theta^+\theta^-\bar\theta^+\eta^-
+\theta^+\theta^-\bar\theta^+\bar\theta^-f
 \end{align}
and $\Phi$ as (\ref{pcomp}),
the gauge transformations of the component fields 
for $Q_\mathcal{N}\ne0$ are given by
 \begin{align}
\delta\phi=&-\bar\lambda,\nonumber\\
%%%%%%%%%%%%%%%%
\delta\bar\psi_-=&-\bar\eta^+-\nabla^{++}\bar\epsilon_+
+(\nabla^3+q)\bar\epsilon_-,\nonumber\\
%%%%%%%%%%%%%%%%
\delta\psi_+=&0,\nonumber\\
%%%%%%%%%%%%%%%%
\delta\bar\psi_+=&\bar\eta^-+\nabla^{--}\bar\epsilon_-
+(\nabla^3-q)\bar\epsilon_+,\nonumber\\
%%%%%%%%%%%%%%%%
\delta\psi_-=&0,\nonumber\\
%%%%%%%%%%%%%%%%
\delta v_{++}=&-\nabla^{--}\bar\lambda,\nonumber\\
%%%%%%%%%%%%%%%%
\delta v_3=&-\nabla^3\bar\lambda,\nonumber\\
%%%%%%%%%%%%%%%%
\delta v_{--}=&-\nabla^{++}\bar\lambda,\nonumber\\
%%%%%%%%%%%%%%%%
\delta v_Y=&-q\bar\lambda,\nonumber\\
%%%%%%%%%%%%%%%%
\delta\varphi=&0,\nonumber\\
%%%%%%%%%%%%%%%%
\delta\omega=&
-\bar f
+\nabla^{++}\bar\lambda_{++}
-\nabla^{--}\bar\lambda_{--}
+\left(\nabla^3-q-1\right)\bar\lambda_{-+}
\nonumber\\
&\hspace{10mm}
+\left(\nabla^3+q+1\right)\bar\lambda_{+-}
+({\cal K}_{(m+1)}+\Delta_\mathcal{N})\bar\tau,\nonumber\\
%%%%%%%%%%%%%%%%
\delta\chi^-=&0,\nonumber\\
%%%%%%%%%%%%%%%%
\delta\bar\chi^+=&
-\nabla^{++}\bar\eta^-
-\left(\nabla^3+p-q-1\right)\bar\eta^+
-({\cal K}_{(m)}+\Delta_\mathcal{N})\bar\epsilon_-,\nonumber\\
%%%%%%%%%%%%%%%%
\delta\chi^+=&0,\nonumber\\
%%%%%%%%%%%%%%%%
\delta\bar\chi^-=&
\nabla^{--}\bar\eta^+
+\left(\nabla^3+q+1\right)\bar\eta^-
+({\cal K}_{(m+1)}+\Delta_\mathcal{N})\bar\epsilon_+,\nonumber\\
%%%%%%%%%%%%%%%%
\delta d=& ({\cal K}_{(m)}+\Delta_\mathcal{N})\bar\lambda.  
 \end{align}
We can fix these invariances by taking the gauge conditions
\begin{equation}
\phi=\omega=\bar\psi_+=\bar\psi_-=0.\label{gcmassive}
\end{equation}
The off-shell supermultiplet 
$\boldsymbol{V}_M(m)=
(v_{++}(m-1),v_3(m),v_{--}(m+1),v_Y(m),\varphi(m+1),d(m);\\
\psi_+(m),\psi_-(m+1),\chi^+(m+1),\chi^-(m),
\bar\chi^+(m),\bar\chi^-(m-1))$
includes six degrees of freedom for both bosons and fermions.
The equations of motion (\ref{eoml0}) therefore reduce to
 \begin{align}
&
d(m)=\varphi(m+1)=0,\nonumber\\
%%%%%%%%%%%%%%%%%%%%
&
({\cal K}_{(m)}+\Delta_\mathcal{N})
\left(
\begin{array}{c}
 v_{++}(m-1)\\
 v_3(m)\\
 v_{--}(m+1)\\
 v_Y(m)\\
\end{array}
\right)=0,\nonumber\\
%%%%%%%%%%%%%%%%%%%%
&
(m-1+j)v_{++}(m-1)-2\left(m-\frac{k}{2}p\right)v_3(m)
+(m+1-j)v_{--}(m+1)+2qv_Y(m)=0,\nonumber\\
%%%%%%%%%%%%%%%%%%%%
&
\left(
\begin{array}{cc}
 m-\frac{k}{2}p-q+1&m+j \\
 m+1-j&m-\frac{k}{2}p+q \\
\end{array}
\right)
\left(
\begin{array}{c}
 \chi^+(m+1)\\
 \chi^-(m)\\
\end{array}
\right)
-2k\Delta_\mathcal{N}
\left(
\begin{array}{c}
 \psi_-(m+1)\\
 \psi_+(m)\\
\end{array}
\right)=0,\nonumber\\
%%%%%%%%%%%%%%%%%%%%
&
\left(
\begin{array}{cc}
 m-\frac{k}{2}p+q& -(m+j) \\
 -(m+1-j)&m-\frac{k}{2}p-q+1 \\
\end{array}
\right)
\left(
\begin{array}{c}
 \psi_-(m+1)\\
 \psi_+(m)\\
\end{array}
\right)
-\left(
\begin{array}{c}
 \chi^+(m+1)\\
 \chi^-(m)\\
\end{array}
\right)=0,\nonumber\\
%%%%%%%%%%%%%%%%%%%%
&
\left(
\begin{array}{cc}
m-\frac{k}{2}p+q &m-1+j \\
m-j &m-\frac{k}{2}p-q-1 \\
\end{array}
\right)
\left(
\begin{array}{c}
 \bar\chi^+(m)\\
 \bar\chi^-(m-1)\\
\end{array}
\right)
=0.\label{eommassive}
 \end{align}
These are the equations of motion for the {\it massive vector} 
supermultiplet on $AdS_3\times S^1$.

The space-time $R$-charge of the vector field
$(v_{++}(m-1),v_3(m),v_{--}(m+1),v_Y(m))$ is
$Q_{s-t}=s+kp$, which coincides with the spectrum of the second
series (with $s\ne0$) of the space-time chiral primaries 
in Ref.~\citen{HHS}.
The equations of motion (\ref{eommassive}) can be solved
by setting $j=1-m=q+1-\frac{k}{2}p$,\footnote{
The value of $m$ is also not fixed by the equations of motion
for $p=0$.}
which leads the on-shell supermultiplet
$(v_{++}(m-1),v_3(m),\\ v_{--}(m+1);\bar\chi^-(m-1),\chi^-(m),\chi^+(m+1))$
to be the space-time chiral $\Delta_{s-t}=\frac{Q_{s-t}}{2}$.

The gauge invariances (\ref{enl}) for the case
$Q_\mathcal{N}=\Delta_\mathcal{N}=0$
written in the component fields
 \begin{align}
  \delta\phi=&-\lambda_{+-}+\lambda,\nonumber\\
%%%%%%%%%%%%%%%%%%%
\delta\bar\psi_-=& -\bar\eta^+-\nabla^{++}\bar\epsilon_+
+(\nabla^3+q)\bar\epsilon_-,\nonumber\\
%%%%%%%%%%%%%%%%%%%
\delta\psi_+=&-\eta^-,\nonumber\\
%%%%%%%%%%%%%%%%%%%
\delta\bar\psi_+=&\bar\eta^-+\nabla^{--}\bar\epsilon_-
-(\nabla^3-q)\bar\epsilon_+,\nonumber\\
%%%%%%%%%%%%%%%%%%%
\delta\psi_-=&\eta^+,\nonumber\\
%%%%%%%%%%%%%%%%%%%
\delta v_{++}=&-\nabla^{--}\bar\lambda,\nonumber\\
%%%%%%%%%%%%%%%%%%%
\delta v_3=&-\nabla^3\bar\lambda,\nonumber\\
%%%%%%%%%%%%%%%%%%%
\delta v_{--}=&-\nabla^{++}\bar\lambda,\nonumber\\
%%%%%%%%%%%%%%%%%%%
\delta v_Y=&-q\bar\lambda,\nonumber\\
%%%%%%%%%%%%%%%%%%%
\delta\omega=&
-\bar f
+\nabla^{++}\bar\lambda_{++}
-\nabla^{--}\bar\lambda_{--}
+(\nabla^3-q-1)\bar\lambda_{-+}\nonumber\\
&\hspace{10mm}
+(\nabla^3+q+1)\bar\lambda_{+-}
+{\cal K}_{(m+1)}\bar\tau,\nonumber\\
%%%%%%%%%%%%%%%%%%%
\delta\varphi=&-f,\nonumber\\
%%%%%%%%%%%%%%%%%%%
\delta\chi^-=&0,\nonumber\\
%%%%%%%%%%%%%%%%%%%
\delta\bar\chi^+=&-\nabla^{++}\bar\eta^--(\nabla^3-q-1)\bar\eta^+
-{\cal K}_{(m)}\bar\epsilon_-,\nonumber\\
%%%%%%%%%%%%%%%%%%%
\delta\chi^+=&0,\nonumber\\
%%%%%%%%%%%%%%%%%%%
\delta\bar\chi^-=&\nabla^{--}\bar\eta^++(\nabla^3+q+1)\bar\eta^-
+{\cal K}_{(m+1)}\bar\epsilon_+,\nonumber\\
%%%%%%%%%%%%%%%%%%%
\delta d=&-{\cal K}_{(m)}\bar\epsilon_-
 \end{align}
can be fixed by taking the gauge
\begin{equation}
\phi=\omega=\varphi=\psi_\pm=\bar\psi_\pm=0.\label{gcmassless}
\end{equation}
The equations of motion (\ref{eoml0}) are given in this
gauge by
 \begin{align}
&
{\cal K}_{(m)}
\left(
\begin{array}{c}
 v_{++}(m-1)\\
 v_3(m)\\
 v_{--}(m+1)\\
 v_Y(m)\\
\end{array}
\right)
+\left(
\begin{array}{c}
 m-j\\
 m-\frac{k}{2}p\\
 m+j\\
 q\\
\end{array}
\right)d(m)=0,\nonumber\\
%%%%%%%%%%%%%%%%%%%%%
  &
d(m)=-\frac{1}{2}\Big((m-1+j)v_{++}(m-1)\nonumber\\
&\hspace{3cm}
+(m+1-j)v_{--}(m+1)-2\left(m-\frac{k}{2}p\right)v_3+2qv_Y(m)\Big),\nonumber\\
%%%%%%%%%%%%%%%%%%%%% 
&
\left(
\begin{array}{cc}
 m+1-kp&m+j \\
 m+1-j&m \\
\end{array}
\right)
\left(
\begin{array}{c}
 \chi^+(m+1)\\
 \chi^-(m)\\
\end{array}
\right)=0,\nonumber\\
%%%%%%%%%%%%%%%%%%%%% 
&
\left(
\begin{array}{cc}
 m&m-1+j \\
 m-j&m-1-kp \\
\end{array}
\right)
\left(
\begin{array}{c}
 \bar\chi^+(m)\\
 \bar\chi^-(m-1)\\
\end{array}
\right)=0,\label{eommassless}
 \end{align}
which are still invariant under the residual
gauge transformation
 \begin{equation}
   \delta
\left(
\begin{array}{c}
 v_{++}(m-1)\\
 v_3(m)\\
 v_{--}(m+1)\\
 v_Y(m)\\
\end{array}
\right)=-
\left(
\begin{array}{c}
 m-j \\
 m-\frac{k}{2}p\\
 m+j \\
 q\\
\end{array}
\right)\bar\lambda(m).\label{rgauge}
 \end{equation}
The equations (\ref{eommassless}) thus should be
interpreted as the Maxwell and Dirac equations
on $AdS_3\times S^1$. The off-shell superfield
$\boldsymbol{V}(m)=(v_{++}(m-1),v_3(m),v_{--}(m+1),v_Y(m),d(m);\\
\bar\chi^-(m-1),\bar\chi^+(m),\chi^-(m),\chi^+(m+1))$
represents the {\it massless vector} supermultiplet.
The on-shell physical fields $(v_{++}(m-1),v_{--}(m+1); \bar\chi^-(m-1),
\chi^+(m+1))$ with $j=1-m=1$ provide the missing state
$s=0$ in the second series.
It should be stressed that the physical spectra differ significantly
in the cases $s=0$ and $s\ne0$. 
The former (latter) is described by the massive (massless)
vector supermultiplet with $3+3$ $(2+2)$ on-shell degrees of freedom
for bosons and fermions.

\section{Space-time superconformal symmetry}\label{stsusy}

To this point we have investigated the hybrid formalism of the
superstring on $AdS_3\times S^1$, preserving
the manifest space-time $N=2$ superconformal symmetry. 
We have also identified the two series
of the space-time chiral primaries and clarified
their supersymmetry structure. They are described 
by the chiral and the vector supermultiplets, but
the structure of the vector supermultiplet depends on
whether it is massive or massless, as is well known. 
The massless vector supermultiplet has fewer
on-shell physical degrees of freedom due to 
the gauge invariance. 

In this section, we present explicit forms of the supersymmetry 
transformations on these supermultiplets by taking 
the Wess-Zumino (WZ)-like gauges (\ref{gcchiral}), (\ref{gcmassive}) 
and (\ref{gcmassless}), 
for which the supersymmetry structure is easy to understand,
because the numbers of fields needed to realize off-shell 
supersymmetry are minimal. We must distinguish two cases here, 
whether $p=0$ or $p\ne0$, because the manners in which
the off-shell supersymmetry is realized differ significantly 
in these cases. 
For vanishing light-cone momentum $p=0$, 
all of the infinite numbers of supersymmetries are realized on
the supermultiplet. For non-vanishing light-cone momentum $p\ne0$, 
by contrast, the space-time symmetry becomes a spectrum 
generating symmetry, since the spectral flow operation 
does not commute with the world-sheet Hamiltonian $L_0$.
Only two of the supersymmetries are closed on 
the supermultiplet, and the others generate 
new physical states with different masses.

Let us begin by investigating the first series, $l=-1$,
found to form the chiral supermultiplet 
$\boldsymbol{\Phi}(m)=(\varphi(m),d(m-1);\chi^+(m),\chi^-(m-1))$
in the WZ-like gauge (\ref{gcchiral}).
For $p=0$, the entire $N=2$ superconformal 
symmetry is realized on the supermultiplet, and 
the component fields are transformed as
 \begin{align}
  \delta^+_r\varphi(m)=&0,\nonumber\\
%%%%%%%%%%%%%%%%%%%%%%
\delta^+_rd(m-1)=&
-\left(m-1-2r+\left(\frac{1}{2}+r\right)j
+\left(\frac{1}{2}-r\right)q\right)\chi^-(m-r-3/2)
\nonumber\\
&\hspace{20mm}
-\left(m-1-2r-\left(\frac{1}{2}-r\right)j-\left(\frac{1}{2}+r\right)q
\right)\chi^+(m-r-1/2),\nonumber\\
%%%%%%%%%%%%%%%%%%%%%%
\delta^+_r\chi^+(m)=&-\left(m-r-\frac{1}{2}+\left(\frac{1}{2}
+r\right)j+\left(\frac{1}{2}-r\right)q\right)
\varphi(m-r-1/2),\nonumber\\
%%%%%%%%%%%%%%%%%%%%%%
\delta^+_r\chi^-(m-1)=&\left(m-r-\frac{1}{2}-\left(\frac{1}{2}
-r\right)j-\left(\frac{1}{2}+r\right)q\right)
\varphi(m-r-1/2),\nonumber\\ 
%%%%%%%%%%%%%%%%%%%%%%%%%%%%%%%%%%%%%%%%%%
\delta^-_r\varphi(m)=&
-\left(\frac{1}{2}-r\right)\chi^-(m-r-1/2)
+\left(\frac{1}{2}+r\right)\chi^+(m-r+1/2),\nonumber\\
%%%%%%%%%%%%%%%%%%%%%%%
\delta^-_rd(m-1)=&0,\nonumber\\
%%%%%%%%%%%%%%%%%%%%%%%
\delta^-_r\chi^+(m)=&\left(\frac{1}{2}-r\right)
d(m-r-1/2),\nonumber\\
%%%%%%%%%%%%%%%%%%%%%%%
\delta^-_r\chi^-(m-1)=&\left(\frac{1}{2}+r\right)
d(m-r-1/2).\label{susytf1}
 \end{align}
These relations are induced from the actions of $\mathcal{G}^\pm_r$, 
(\ref{stgp}) and (\ref{stgm}). 
We can confirm that 
they actually satisfy the $N=2$ superconformal algebra 
with vanishing central charge $c=6kp=0$:
 \begin{alignat}{2}
\left[\delta^\mathcal{L}_l,\delta^\mathcal{L}_n\right]=&
-(l-n)\delta^\mathcal{L}_{n+l},&\qquad
%%%%%
\left[\delta^\mathcal{L}_n,\delta^\pm_r\right]=&
-\left(\frac{n}{2}-r\right)\delta^\pm_{n+r},\nonumber\\
%%%%%%%%%%%%%%%%%%
\left[\delta^\mathcal{L}_l,\delta^\mathcal{I}_n\right]=&
n\delta^\mathcal{I}_{n+l},&\qquad
%%%%%
\left[\delta^\mathcal{I}_n,\delta^\pm_r\right]=&
\mp\delta^\pm_{n+r},\nonumber\\
%%%%%%%%%%%%%%%%%%
  \left\{\delta^+_r,\delta^-_s\right\}=&
\delta^\mathcal{L}_{r+s}+\frac{1}{2}(r-s)\delta^\mathcal{I}_{r+s},& &
\label{sca}
 \end{alignat}
where $\delta^\mathcal{L}_n$ and $\delta^\mathcal{I}_n$ are bosonic
transformations induced from the actions of 
$\mathcal{L}_n$ (\ref{stvir}) and $\mathcal{I}_n$ (\ref{stu1}), 
whose explicit forms are given in Appendix C.

For $p\ne0$, however, only two supersymmetries,
$\delta^\pm_{\mp\frac{1}{2}}$, are closed on the chiral 
supermultiplet:
 \begin{align}
\delta^+_{-\frac{1}{2}}\varphi(m)=&0,\nonumber\\
%%%%%%%%%%%%%%%%%
\delta^+_{-\frac{1}{2}}d(m-1)=&
-(m-j)\chi^+(m)-(m-\frac{k}{2}p+q)\chi^-(m-1),\nonumber\\
%%%%%%%%%%%%%%%%%
\delta^+_{-\frac{1}{2}}\chi^+(m)=&
-(m-\frac{k}{2}p+q)\varphi(m),\nonumber\\
%%%%%%%%%%%%%%%%%
\delta^+_{-\frac{1}{2}}\chi^-(m-1)=&
(m-j)\varphi(m),\nonumber\\
%%%%%%%%%%%%%%%%%
\delta^-_{\frac{1}{2}}\varphi(m)=&\chi^+(m),\nonumber\\
%%%%%%%%%%%%%%%%%
\delta^-_{\frac{1}{2}}d(m-1)=&0,\nonumber\\
%%%%%%%%%%%%%%%%%
\delta^-_{\frac{1}{2}}\chi^+(m)=&0,\nonumber\\
%%%%%%%%%%%%%%%%%
\delta^-_{\frac{1}{2}}\chi^-(m-1)=&d(m-1),\nonumber\\
%%%%%%%%%%%%%%%%%
\left(\delta^\mathcal{L}_0-\frac{1}{2}\delta^\mathcal{I}_0\right)
\boldsymbol{\Phi}(m)=&
-\left(m-\frac{k}{2}p+q\right)\boldsymbol{\Phi}(m).
 \end{align}
These satisfy the three-dimensional (one bosonic and two fermionic)
subalgebra of (\ref{sca}) defined by the single non-trivial relation
\begin{equation}
 \left\{\delta^+_{-\frac{1}{2}},\delta^-_{\frac{1}{2}}\right\}=
\delta^\mathcal{L}_0-\frac{1}{2}\delta^\mathcal{I}_0.\label{subsca}
\end{equation}
The right-hand side vanishes on shell, because
the on-shell fields are space-time chiral primaries.

The two cases for $l=0$, {\it i.e.} $Q_\mathcal{N}\ne0$ and $Q_\mathcal{N}=0$,
have similar structures.
For $p=0$ in the $Q_\mathcal{N}\ne0$ case,
the superconformal transformations on the massive vector 
supermultiplet
$\boldsymbol{V}_M(m)=
(v_{++}(m-1),v_3(m),v_{--}(m+1),v_Y(m),\varphi(m+1),d(m);
\psi_+(m),\psi_-(m+1),\\ 
\chi^+(m+1),\chi^-(m),\bar\chi^+(m),\bar\chi^-(m-1))$
are given by
 \begin{align}
\delta^+_rv_{++}(m-1)=&
-\left(\frac{1}{2}+r\right)\chi^-(m-r-1/2)\nonumber\\
&
-\left(\frac{1}{2}-r\right)(m-j)\psi_+(m-r-1/2)
-\left(\frac{1}{2}+r\right)(m-j)\psi_-(m-r+1/2),\nonumber\\
%%%%%%%%%%%%%%%%%%%%%% 
\delta^+_rv_3(m)=&
 \frac{1}{2}\left(\frac{1}{2}-r\right)\chi^-(m-r-1/2)
+\frac{1}{2}\left(\frac{1}{2}+r\right)\chi^+(m-r+1/2)
\nonumber\\
&
-\left(\frac{1}{2}-r\right)m\psi_+(m-r-1/2)
-\left(\frac{1}{2}+r\right)m\psi_-(m-r+1/2),\nonumber\\
%%%%%%%%%%%%%%%%%%%%%% 
\delta^+_rv_{--}(m+1)=&
-\left(\frac{1}{2}-r\right)\chi^+(m-r+1/2)\nonumber\\
&
-\left(\frac{1}{2}-r\right)(m+j)\psi_+(m-r-1/2)
-\left(\frac{1}{2}+r\right)(m+j)\psi_-(m-r+1/2),\nonumber\\
%%%%%%%%%%%%%%%%%%%%%% 
\delta^+_rv_Y(m)=&
-\frac{1}{2}\left(\frac{1}{2}-r\right)\chi^-(m-r-1/2)
+\frac{1}{2}\left(\frac{1}{2}+r\right)\chi^+(m-r+1/2)\nonumber\\
&
-\left(\frac{1}{2}-r\right)q\psi_+(m-r-1/2)
-\left(\frac{1}{2}+r\right)q\psi_-(m-r+1/2),\nonumber\\
%%%%%%%%%%%%%%%%%%%%%% 
\delta^+_r\varphi(m+1)=&0,\nonumber\\
%%%%%%%%%%%%%%%%%%%%%% 
\delta^+_r d(m)=&\left(\frac{1}{2}-r\right)\left(-j(j-1)+
q(q+1)\right)\psi_+(m-r-1/2)\nonumber\\
&
+\left(\frac{1}{2}+r\right)\left(-j(j-1)+
q(q+1)\right)\psi_-(m-r+1/2),\nonumber\\
%%%%%%%%%%%%%%%%%%%%%% 
  \delta^+_r\psi_+(m)=&
-\left(\frac{1}{2}+r\right)\varphi(m-r+1/2),\nonumber\\
%%%%%%%%%%%%%%%%%%%%%% 
\delta^+_r\psi_-(m+1)=&
\left(\frac{1}{2}-r\right)\varphi(m-r+1/2),\nonumber\\
%%%%%%%%%%%%%%%%%%%%%% 
\delta^+_r\chi^+(m+1)=&0,\nonumber\\
%%%%%%%%%%%%%%%%%%%%%% 
\delta^+_r\chi^-(m)=&0,\nonumber\\
%%%%%%%%%%%%%%%%%%%%%% 
\delta^+_r\bar\chi^+(m)=&-\left(\frac{1}{2}-r\right)d(m-r-1/2)\nonumber\\
&
-\left(\frac{1}{2}-r\right)(m-1+j)v_{++}(m-r-3/2)\nonumber\\
&
-\left(\frac{1}{2}+r\right)(m-1+j)(v_3-v_Y)(m-r-1/2)\nonumber\\
&
+\left(\frac{1}{2}-r\right)(m-1-q)(v_3+v_Y)(m-r-1/2)\nonumber\\
&
+\left(\frac{1}{2}+r\right)(m-1-q)v_{--}(m-r+1/2),\nonumber\\
%%%%%%%%%%%%%%%%%%%%%% 
\delta^+_r\bar\chi^-(m-1)=&-\left(\frac{1}{2}+r\right)d(m-r-1/2)\nonumber\\
&
+\left(\frac{1}{2}-r\right)(m+q)v_{++}(m-r-3/2)\nonumber\\
&
+\left(\frac{1}{2}+r\right)(m+q)(v_3-v_Y)(m-r-1/2)\nonumber\\
&
-\left(\frac{1}{2}-r\right)(m-j)(v_3+v_Y)(m-r-1/2)\nonumber\\
&
-\left(\frac{1}{2}+r\right)(m-j)v_{--}(m-r+1/2),\nonumber\\
%%%%%%%%%%%%%%%%%%%%%% 
\delta^-_rv_{++}(m-1)=&
-\left(\frac{1}{2}+r\right)\bar\chi^-(m-r-1/2),\nonumber\\
%%%%%%%%%%%%%%%%%%%%%% 
\delta^-_rv_3(m)=&
\left(\frac{1}{2}+r\right)\bar\chi^+(m-r+1/2)
+\frac{1}{2}\left(\frac{1}{2}-r\right)\bar\chi^-(m-r-1/2),\nonumber\\
%%%%%%%%%%%%%%%%%%%%%% 
\delta^-_rv_{--}(m+1)=&
-\left(\frac{1}{2}-r\right)\bar\chi^+(m-r+1/2),\nonumber\\
%%%%%%%%%%%%%%%%%%%%%% 
\delta^-_rv_Y(m)=&
-\left(\frac{1}{2}+r\right)\bar\chi^+(m-r+1/2)
+\frac{1}{2}\left(\frac{1}{2}-r\right)\bar\chi^-(m-r-1/2),\nonumber\\
%%%%%%%%%%%%%%%%%%%%%% 
\delta^-_r\varphi(m+1)=&
-\left(\frac{1}{2}-r\right)\chi^-(m-r+1/2)
+\left(\frac{1}{2}+r\right)\chi^+(m-r+3/2)\nonumber\\
&
+\Bigg(\left(\frac{1}{2}+r\right)\left(m-r+\frac{1}{2}+j\right)\nonumber\\
&\hspace{3cm}
+\left(\frac{1}{2}-r\right)\left(m-r+\frac{3}{2}-q\right)\Bigg)
\psi_+(m-r+1/2)\nonumber\\
&
-\Bigg(\left(\frac{1}{2}-r\right)\left(m-r+\frac{3}{2}-j\right)\nonumber\\
&\hspace{3cm}
+\left(\frac{1}{2}+r\right)\left(m-r+\frac{1}{2}+q\right)\Bigg)
\psi_-(m-r+3/2),\nonumber\\
%%%%%%%%%%%%%%%%%%%%%% 
\delta^-_r d(m)=&\Bigg(\left(\frac{1}{2}-r\right)
\left(m-r+\frac{1}{2}-j\right)\nonumber\\
&\hspace{3cm}
+\left(\frac{1}{2}+r\right)
\left(m-r+\frac{1}{2}+q\right)\Bigg)\bar\chi^+(m-r+1/2)\nonumber\\
&
+\Bigg(\left(\frac{1}{2}+r\right)
\left(m-r-\frac{1}{2}+j\right)\nonumber\\
&\hspace{3cm}
+\left(\frac{1}{2}-r\right)
\left(m-r-\frac{1}{2}-q\right)\Bigg)\bar\chi^-(m-r-1/2),\nonumber\\
%%%%%%%%%%%%%%%%%%%%%% 
\delta^-_r\psi_+(m)=&
\left(\frac{1}{2}-r\right)v_{++}(m-r-1/2)
-\left(\frac{1}{2}+r\right)(v_3+v_Y)(m-r+1/2),\nonumber\\
%%%%%%%%%%%%%%%%%%%%%% 
\delta^-_r\psi_-(m+1)=&
\left(\frac{1}{2}+r\right)v_{--}(m-r+3/2)
+\left(\frac{1}{2}-r\right)(v_3-v_Y)(m-r+1/2),\nonumber\\
%%%%%%%%%%%%%%%%%%%%%% 
\delta^-_r\chi^+(m+1)=&\left(\frac{1}{2}-r\right)d(m-r+1/2)
\nonumber\\
&
+\Bigg(\left(\frac{1}{2}-r\right)\left(m-r+\frac{3}{2}-j\right)\nonumber\\
&\hspace{3cm}
+\left(\frac{1}{2}+r\right)\left(m+q\right)\Bigg)
v_{--}(m-r+3/2)\nonumber\\
&
-\Bigg(\left(\frac{1}{2}+r\right)\left(m-r+\frac{1}{2}+j\right)\nonumber\\
&\hspace{3cm}
+\left(\frac{1}{2}-r\right)\left(m-2r-q\right)\Bigg)
(v_3+v_Y)(m-r+1/2)\nonumber\\
&
-\left(\frac{1}{2}-r\right)^2(v_3-v_Y)(m-r+1/2)
-\left(\frac{1}{4}-r^2\right)v_{++}(m-r-1/2),\nonumber\\
%%%%%%%%%%%%%%%%%%%%%% 
\delta^-_r\chi^-(m)=&\left(\frac{1}{2}+r\right)d(m-r+1/2)
\nonumber\\
&
+\Bigg(\left(\frac{1}{2}+r\right)\left(m-r-\frac{1}{2}+j\right)\nonumber\\
&\hspace{3cm}
+\left(\frac{1}{2}-r\right)\left(m+1-q\right)\Bigg)
v_{++}(m-r-1/2)\nonumber\\
&
-\Bigg(\left(\frac{1}{2}-r\right)\left(m-r+\frac{1}{2}-j\right)\nonumber\\
&\hspace{3cm}
+\left(\frac{1}{2}+r\right)\left(m-2r+1+q\right)\Bigg)
(v_3-v_Y)(m-r+1/2)\nonumber\\
&
+\left(\frac{1}{2}+r\right)^2(v_3+v_Y)(m-r+1/2)
+\left(\frac{1}{4}-r^2\right)v_{--}(m-r+3/2),\nonumber\\
%%%%%%%%%%%%%%%%%%%%%% 
\delta^-_r\bar\chi^+(m)=&0,\nonumber\\
%%%%%%%%%%%%%%%%%%%%%% 
\delta^-_r\bar\chi^-(m-1)=&0.\label{susytf2}
\end{align}
These also satisfy the $N=2$ superconformal algebra (\ref{sca}).
In this case, the two supersymmetry transformations on the $p\ne0$ 
fields become
 \begin{align}
 \delta^+_{-\frac{1}{2}}v_{++}(m-1)=&-(m-j)\psi_+(m),\nonumber\\
%%%%%%%%%%%%%%%%%%%%%%%
 \delta^+_{-\frac{1}{2}}v_3(m)=&\frac{1}{2}\chi^-(m)-(m-\frac{k}{2}p)\psi_+(m),\nonumber\\
%%%%%%%%%%%%%%%%%%%%%%%
 \delta^+_{-\frac{1}{2}}v_{--}(m+1)=&-\chi^+(m+1)-(m+j)\psi_+(m),\nonumber\\
%%%%%%%%%%%%%%%%%%%%%%%
 \delta^+_{-\frac{1}{2}}v_Y(m)=&-\frac{1}{2}\chi^-(m)-q\psi_+(m),\nonumber\\
%%%%%%%%%%%%%%%%%%%%%%%
 \delta^+_{-\frac{1}{2}}\varphi(m+1)=&0,\nonumber\\
%%%%%%%%%%%%%%%%%%%%%%%
 \delta^+_{-\frac{1}{2}}d(m)=&
k{\cal K}_{(m)}\psi_+(m),\nonumber\\
%%%%%%%%%%%%%%%%%%%%%%%
   \delta^+_{-\frac{1}{2}}\psi_+(m)=&0,\nonumber\\
%%%%%%%%%%%%%%%%%%%%%%%
 \delta^+_{-\frac{1}{2}}\psi_-(m+1)=&\varphi(m+1),\nonumber\\
%%%%%%%%%%%%%%%%%%%%%%%
 \delta^+_{-\frac{1}{2}}\chi^+(m+1)=&0,\nonumber\\
%%%%%%%%%%%%%%%%%%%%%%%
 \delta^+_{-\frac{1}{2}}\chi^-(m)=&0,\nonumber\\
%%%%%%%%%%%%%%%%%%%%%%%
 \delta^+_{-\frac{1}{2}}\bar\chi^+(m)=&-d(m)-(m-1+j)v_{++}(m-1)\nonumber\\
&\hspace{3cm}
+\left(m-\frac{k}{2}p-q-1\right)(v_3+v_Y)(m),\nonumber\\
%%%%%%%%%%%%%%%%%%%%%%%
 \delta^+_{-\frac{1}{2}}\bar\chi^-(m-1)=&
\left(m-\frac{k}{2}p+q\right)v_{++}(m-1)-(m-j)(v_3+v_Y)(m-1),\nonumber\\
%%%%%%%%%%%%%%%%%%%%%%%
 \delta^-_{\frac{1}{2}}v_{++}(m-1)=&-\bar\chi^-(m-1),\nonumber\\
%%%%%%%%%%%%%%%%%%%%%%%
 \delta^-_{\frac{1}{2}}v_3(m)=&\frac{1}{2}\bar\chi^+(m),\nonumber\\
%%%%%%%%%%%%%%%%%%%%%%%
 \delta^-_{\frac{1}{2}}v_{--}(m+1)=&0,\nonumber\\
%%%%%%%%%%%%%%%%%%%%%%%
 \delta^-_{\frac{1}{2}}v_Y(m)=&-\frac{1}{2}\bar\chi^+(m),\nonumber\\
%%%%%%%%%%%%%%%%%%%%%%%
 \delta^-_{\frac{1}{2}}\varphi(m+1)=&\chi^+(m+1)+(m+j)\psi_+(m)
-\left(m-\frac{k}{2}p+q\right)\psi_-(m+1),\nonumber\\
%%%%%%%%%%%%%%%%%%%%%%%
 \delta^-_{\frac{1}{2}}d(m)=&(m-1+j)\bar\chi^-(m-1)
+\left(m-\frac{k}{2}p+q\right)\bar\chi^+(m),\nonumber\\
%%%%%%%%%%%%%%%%%%%%%%%
 \delta^-_{\frac{1}{2}}\psi_+(m)=&(v_3+v_Y)(m),\nonumber\\
%%%%%%%%%%%%%%%%%%%%%%%
 \delta^-_{\frac{1}{2}}\psi_-(m+1)=&v_{--}(m+1),\nonumber\\
%%%%%%%%%%%%%%%%%%%%%%%
 \delta^-_{\frac{1}{2}}\chi^+(m+1)=&-(m+j)(v_3+v_Y)(m)
+\left(m-\frac{k}{2}p+q\right)v_{--}(m+1),\nonumber\\
%%%%%%%%%%%%%%%%%%%%%%%
 \delta^-_{\frac{1}{2}}\chi^-(m)=&d(m)+(m-1+j)v_{++}(m-1)\nonumber\\
&-\left(m-\frac{k}{2}p+q\right)(v_3-v_Y)(m)+(v_3+v_Y)(m),\nonumber\\
%%%%%%%%%%%%%%%%%%%%%%%
 \delta^-_{\frac{1}{2}}\bar\chi^+(m)=&0,\nonumber\\
%%%%%%%%%%%%%%%%%%%%%%%
 \delta^-_{\frac{1}{2}}\bar\chi^-(m-1)=&0,\nonumber\\
%%%%%%%%%%%%%%%%%%%%%%%
 \left(\delta^\mathcal{L}_0-\frac{1}{2}\delta^\mathcal{I}_0\right)
\boldsymbol{V}_M(m)
=&-\left(m-\frac{k}{2}p+q\right)
\boldsymbol{V}_M(m)
v_{++}(m-1).
 \end{align}
This satisfies the subalgebra (\ref{subsca}),
whose right-hand side vanishes on shell also in this case.

In the case of a massless vector supermultiplet, 
the superconformal transformations for $p=0$ are given by
 \begin{align}
   \delta^+_rv_{++}(m-1)=&
-\left(\frac{1}{2}+r\right)\chi^-(m-r-1/2),\nonumber\\
%%%%%%%%%%%%%%%%%%%%%
 \delta^+_rv_3(m)=&
\frac{1}{2}\left(\frac{1}{2}-r\right)\chi^-(m-r-1/2)
+\frac{1}{2}\left(\frac{1}{2}+r\right)\chi^+(m-r+1/2),\nonumber\\
%%%%%%%%%%%%%%%%%%%%%
 \delta^+_rv_{--}(m+1)=&
-\left(\frac{1}{2}-r\right)\chi^+(m-r+1/2),\nonumber\\
%%%%%%%%%%%%%%%%%%%%%
 \delta^+_rv_Y(m)=&
-\frac{1}{2}\left(\frac{1}{2}-r\right)\chi^-(m-r-1/2)
+\frac{1}{2}\left(\frac{1}{2}+r\right)\chi^+(m-r+1/2),\nonumber\\
%%%%%%%%%%%%%%%%%%%%%
\delta^+_rd(m)=&0,\nonumber\\
%%%%%%%%%%%%%%%%%%%%%
\delta^+_r\chi^+(m+1)=&0,\nonumber\\
%%%%%%%%%%%%%%%%%%%%%
\delta^+_r\chi^-(m)=&0,\nonumber\\
%%%%%%%%%%%%%%%%%%%%%
\delta^+_r\bar\chi^+(m)=&-\left(\frac{1}{2}-r\right)d(m-r-1/2)
\nonumber\\
&
-\left(\frac{1}{2}-r\right)(m-1+j)v_{++}(m-r-3/2)\nonumber\\
&
-\left(\frac{1}{2}+r\right)(m-1+j)(v_3-v_Y)(m-r-1/2)
\nonumber\\
&
+\left(\frac{1}{2}-r\right)(m-1)(v_3+v_Y)(m-r-1/2)\nonumber\\
&
+\left(\frac{1}{2}+r\right)(m-1)v_{--}(m-r+1/2),\nonumber\\
%%%%%%%%%%%%%%%%%%%%%
\delta^+_r\bar\chi^-(m-1)=&-\left(\frac{1}{2}+r\right)d(m-r-1/2)
\nonumber\\
&
+\left(\frac{1}{2}-r\right)mv_{++}(m-r-3/2)\nonumber\\
&
+\left(\frac{1}{2}+r\right)m(v_3-v_Y)(m-r-1/2)\nonumber\\
&
-\left(\frac{1}{2}-r\right)(m-j)(v_3+v_Y)(m-r-1/2)\nonumber\\
&
-\left(\frac{1}{2}+r\right)(m-j)v_{--}(m-r+1/2),\nonumber\\
%%%%%%%%%%%%%%%%%%%%%
\delta^-_rv_{++}(m-1)=&
-\left(\frac{1}{2}+r\right)\bar\chi^-(m-r-1/2),\nonumber\\
%%%%%%%%%%%%%%%%%%%%%
\delta^-_rv_3(m)=&
\frac{1}{2}\left(\frac{1}{2}+r\right)\bar\chi^+(m-r+1/2)
+\frac{1}{2}\left(\frac{1}{2}-r\right)\bar\chi^-(m-r-1/2),\nonumber\\
%%%%%%%%%%%%%%%%%%%%%
\delta^-_rv_{--}(m+1)=&
-\left(\frac{1}{2}-r\right)\bar\chi^+(m-r+1/2),\nonumber\\
%%%%%%%%%%%%%%%%%%%%%
\delta^-_rv_Y(m)=&
-\frac{1}{2}\left(\frac{1}{2}+r\right)\bar\chi^+(m-r+1/2)
+\frac{1}{2}\left(\frac{1}{2}-r\right)\bar\chi^-(m-r-1/2),\nonumber\\
%%%%%%%%%%%%%%%%%%%%%
\delta^-_rd(m)=&
\left(m-\left(\frac{1}{2}-r\right)(j-1)\right)\bar\chi^+(m-r+1/2)
\nonumber\\
&
+\left(m+\left(\frac{1}{2}+r\right)(j-1)\right)\bar\chi^-(m-r-1/2),\nonumber\\
%%%%%%%%%%%%%%%%%%%%%
\delta^-_r\chi^+(m+1)=&\left(\frac{1}{2}-r\right)d(m-r+1/2)
\nonumber\\
&
-\Bigg(\left(\frac{1}{2}+r\right)\left(m-r+\frac{1}{2}+j\right)\nonumber\\
&\hspace{3cm}
+\left(\frac{1}{2}-r\right)(m-2r)\Bigg)(v_3+v_Y)(m-r+1/2)
\nonumber\\
&
+\Bigg(\left(\frac{1}{2}-r\right)\left(m-r+\frac{3}{2}-j\right)\nonumber\\
&\hspace{3cm}
+\left(\frac{1}{2}+r\right)m\Bigg)v_{--}(m-r+3/2)
\nonumber\\
&
-\left(\frac{1}{2}-r\right)^2(v_3-v_Y)(m-r+1/2)
-\left(\frac{1}{4}-r^2\right)v_{++}(m-r-1/2),\nonumber\\
%%%%%%%%%%%%%%%%%%%%%
\delta^-_r\chi^-(m)=&\left(\frac{1}{2}+r\right)d(m-r+1/2)
\nonumber\\
&
+\Bigg(\left(\frac{1}{2}+r\right)\left(m-r-\frac{1}{2}+j\right)\nonumber\\
&\hspace{3cm}
+\left(\frac{1}{2}-r\right)(m+1)\Bigg)v_{++}(m-r-1/2)
\nonumber\\
&
-\Bigg(\left(\frac{1}{2}-r\right)\left(m-r+\frac{1}{2}-j\right)\nonumber\\
&\hspace{3cm}
+\left(\frac{1}{2}+r\right)(m+1-2r)\Bigg)(v_3-v_Y)(m-r+1/2)
\nonumber\\
&
+\left(\frac{1}{2}+r\right)^2(v_3+v_Y)(m-r+1/2)
+\left(\frac{1}{4}-r^2\right)v_{--}(m-r+3/2),\nonumber\\
%%%%%%%%%%%%%%%%%%%%%
\delta^-_r\bar\chi^+(m)=&0,\nonumber\\
%%%%%%%%%%%%%%%%%%%%%
\delta^-_r\bar\chi^-(m-1)=&0,\label{susytf3}
 \end{align}
which satisfy the $N=2$ superconformal algebra (\ref{sca}) 
{\it up to} the gauge transformation (\ref{rgauge}), e.g.
\begin{equation}
 \left\{\delta^+_r,\delta^-_s\right\}=
\delta^\mathcal{L}_{r+s}+\frac{1}{2}(r-s)\delta^\mathcal{I}_{r+s}
+\delta(\bar\lambda).\label{msusy}
\end{equation}
The gauge parameter $\bar\lambda$ is field dependent and given by
\begin{align}
 \bar\lambda(m)=&
-\left(\frac{1}{2}-r\right)\left(\frac{1}{2}+s\right)(v_3+v_Y)(m-r-s) 
\nonumber\\
&
-\left(\frac{1}{2}+r\right)\left(\frac{1}{2}+s\right)v_{--}(m-r-s+1) 
\nonumber\\
&
-\left(\frac{1}{2}-r\right)\left(\frac{1}{2}-s\right)v_{++}(m-r-s-1) 
\nonumber\\
&
-\left(\frac{1}{2}+r\right)\left(\frac{1}{2}-s\right)(v_3-v_Y)(m-r-s). 
\end{align}
The two manifest supersymmetries for $p\ne0$ are now
\begin{align}
  \delta^+_{-\frac{1}{2}}v_{++}(m-1)=&0,\nonumber\\
%%%%%%%%%%%%%%%%%%%% 
 \delta^+_{-\frac{1}{2}}v_3(m)=&\frac{1}{2}\chi^-(m),\nonumber\\
%%%%%%%%%%%%%%%%%%%% 
 \delta^+_{-\frac{1}{2}}v_{--}(m+1)=&-\chi^+(m+1),\nonumber\\
%%%%%%%%%%%%%%%%%%%% 
 \delta^+_{-\frac{1}{2}}v_Y(m)=&-\frac{1}{2}\chi^-(m),\nonumber\\
%%%%%%%%%%%%%%%%%%%%
 \delta^+_{-\frac{1}{2}}d(m)=&0,\nonumber\\ 
%%%%%%%%%%%%%%%%%%%% 
 \delta^+_{-\frac{1}{2}}\chi^+(m+1)=&0,\nonumber\\
%%%%%%%%%%%%%%%%%%%% 
 \delta^+_{-\frac{1}{2}}\chi^-(m)=&0,\nonumber\\
%%%%%%%%%%%%%%%%%%%% 
 \delta^+_{-\frac{1}{2}}\bar\chi^+(m)=&-d(m)-(m-1+j)v_{++}(m-1)
+\left(m-\frac{k}{2}p-q-1\right)(v_3+v_Y)(m),\nonumber\\
%%%%%%%%%%%%%%%%%%%% 
 \delta^+_{-\frac{1}{2}}\bar\chi^-(m-1)=&-(m-j)(v_3+v_Y)(m)
+\left(m-\frac{k}{2}p+q\right)v_{++}(m-1),\nonumber\\
%%%%%%%%%%%%%%%%%%%% 
 \delta^-_{\frac{1}{2}}v_{++}(m-1)=&-\bar\chi^-(m-1),\nonumber\\
%%%%%%%%%%%%%%%%%%%% 
 \delta^-_{\frac{1}{2}}v_3(m)=&\frac{1}{2}\bar\chi^+(m),\nonumber\\
%%%%%%%%%%%%%%%%%%%% 
 \delta^-_{\frac{1}{2}}v_{--}(m+1)=&0,\nonumber\\
%%%%%%%%%%%%%%%%%%%% 
 \delta^-_{\frac{1}{2}}v_Y(m)=&-\frac{1}{2}\bar\chi^+(m),\nonumber\\
%%%%%%%%%%%%%%%%%%%% 
 \delta^-_{\frac{1}{2}}d(m)=&(m-1+j)\bar\chi^-(m-1)
+\left(m-\frac{k}{2}p+q\right)\bar\chi^+(m),\nonumber\\
%%%%%%%%%%%%%%%%%%%% 
 \delta^-_{\frac{1}{2}}\chi^+(m+1)=&-(m+j)(v_3+v_Y)(m)
+\left(m-\frac{k}{2}p+q\right)v_{--}(m+1),\nonumber\\
%%%%%%%%%%%%%%%%%%%% 
 \delta^-_{\frac{1}{2}}\chi^-(m)=&d(m)+(m-1+j)v_{++}(m-1)\nonumber\\
&
-\left(m-\frac{k}{2}p+q\right)(v_3-v_Y)(m)+(v_3+v_Y)(m),\nonumber\\
%%%%%%%%%%%%%%%%%%%% 
 \delta^-_{\frac{1}{2}}\bar\chi^+(m)=&0,\nonumber\\
%%%%%%%%%%%%%%%%%%%%
 \delta^-_{\frac{1}{2}}\bar\chi^-(m-1)=&0,\nonumber\\ 
%%%%%%%%%%%%%%%%%%%% 
 \left(\delta^\mathcal{L}_0-\frac{1}{2}\delta^\mathcal{I}_0\right)
\boldsymbol{V}(m)=&-m\boldsymbol{V}(m),
\end{align}
where we have used $q=\frac{k}{2}p$, obtained from the $U(1)$ constraint.
They also satisfy the sub-algebra of the modified supersymmetry
(\ref{msusy}),
\begin{equation}
  \left\{\delta^+_{-\frac{1}{2}},\delta^-_{\frac{1}{2}}\right\}=
\delta^\mathcal{L}_0-\frac{1}{2}\delta^\mathcal{I}_0
+\delta(\bar\lambda),
\end{equation}
with the gauge parameter
\begin{equation}
 \bar\lambda(m)=-(v_3+v_Y)(m).
\end{equation}
These two supersymmetries are anti-commutative on shell
up to the gauge transformation.

\section{Summary and discussion}\label{concl}

In this paper, we have studied superstrings on $AdS_3\times S^1$
using a hybrid formalism. The description was obtained
through a field redefinition from the world-sheet fields 
in the RNS formalism.\cite{MO,HHS} We found that 
the space-time supersymmetry is 
manifestly preserved and closed off shell.
The physical spectrum was investigated and 
identified with two series of space-time chiral primaries found 
from the analysis in the RNS formalism\cite{HHS}. 
While the first series is simply represented by an $AdS$ analog 
of the chiral supermultiplet, the second series is described by
different supermultiplets in the two cases $Q_\mathcal{N}=0$ and
$Q_\mathcal{N}\ne0$.
The former (latter) is the massless (massive) vector supermultiplet 
with two (three) on-shell physical degrees of freedom for both bosons 
and fermions.

The supersymmetries on $AdS_3\times S^1$ can be enlarged to 
the boundary $N=2$ superconformal symmetry. 
The entire infinite-dimensional symmetry is realized on the 
vanishing light-cone momentum, $p=0$, states.
This results from 
the fact that the mass spectrum is degenerated 
with respect to the momentum $m$. 
For non-vanishing light-cone momentum, $p\ne0$, however, this degeneracy 
is lifted, and the superconformal symmetry becomes 
the spectrum generating symmetry. Only two of them are closed on 
a supermultiplet. These two form a simple subalgebra
whose right-hand side vanishes on shell. The other symmetries
generate new physical states with different masses.

The Penrose limit of $AdS_3\times S^1$ gives an NS-NS plane wave 
background.\cite{HS} Hybrid superstrings propagating on this background
have already been studied.\cite{K} 
It is interesting to compare the results of the two models and trace, 
for example, the transition of the physical spectra in this limit.
Such an analysis for superstrings on $AdS_3\times S^3$ 
was recently given within the RNS formalism.\cite{JS} 
We can carry out a similar analysis for hybrid superstrings 
on $AdS_3\times S^1$, keeping all the supersymmetries manifest. 
We hope to report the results of such analysis in the future.

\section*{Acknowledgements}
This work is supported in part by 
a Grant-in-Aid for Scientific Research 
(No. 13135213) from the Ministry of Education, Culture,
Sports, Science and Technology, Japan.

\appendix
\section{The Space-Time $N=2$ Superconformal Symmetry and the $sl(1|2)$
 Current Algebra}\label{appA}

In this appendix, we rewrite the space-time
$N=2$ superconformal generators (\ref{scharge}) in terms
of the currents of the global $sl(1|2)$ symmetry.

First, we introduce the local (holomorphic) currents for the global
subalgebra $sl(1|2)$ of the $N=2$ superconformal algebra as
\begin{alignat}{4}
  \mathcal{L}_{\pm1}=&-\oint\dz J^{\pm\pm},&\quad 
%%%%%%%%%%%%%%%%%%%
 \mathcal{L}_0=&-\oint\dz J^3,&\quad
%%%%%%%%%%%%%%%%%%%
 \mathcal{I}_0=&\oint\dz J^Y,\nonumber\\
%%%%%%%%%%%%%%%%%%%
 \mathcal{G}^+_{\pm\frac{1}{2}}=&\oint\dz \bar Q^\pm,&\quad
%%%%%%%%%%%%%%%%%%%
 \mathcal{G}^-_{\pm\frac{1}{2}}=&\oint\dz Q^\pm, & &
\end{alignat}
where
\begin{align}
  J^{\pm\pm}=&-e^{\mp\beta i( X^0+ X^1)}
(\frac{1}{\beta}i\partial X^1\mp\frac{1}{Q}\partial\phi_L)
\pm\Theta^\pm\mathcal{P}_\mp
\pm\bar\Theta^\pm\bar{\mathcal{P}}_\mp,
\nonumber\\
%%%%%%%%%%%%%%%%%%%%%
J_3=&\frac{1}{\beta}i\partial X^0
+\frac{1}{2}(
\Theta^+\mathcal{P}_+
-\Theta^-\mathcal{P}_-
+\bar\Theta^+\bar{\mathcal{P}}_+
-\bar\Theta^-\bar{\mathcal{P}}_-),
\nonumber\\
%%%%%%%%%%%%%%%%%%%%%
J^Y=&2(\frac{1}{Q}i\partial\widehat Y-i\partial\rho)
+(\Theta^+\mathcal{P}_++\Theta^-\mathcal{P}_--\bar\Theta^+\bar{\mathcal{P}}_+
-\bar\Theta^-\bar{\mathcal{P}}_-),
\nonumber\\
%%%%%%%%%%%%%%%%%
  Q^\pm=&
\mathcal{P}_\mp+\bar\Theta^\mp e^{\mp\beta i( X^0+ X^1)}
(\frac{1}{\beta}i\partial X^1\mp\frac{1}{Q}\partial\phi_L)\nonumber\\
&\hspace{10mm}
-\bar\Theta^\pm(\frac{1}{\beta}i\partial X^0\pm
\frac{1}{Q}i\partial\widehat Y\mp i\partial\rho
\pm\Theta^\pm\mathcal{P}_\pm\mp\bar\Theta^\mp\bar{\mathcal{P}}_\mp)\nonumber\\
&\hspace{15mm}
\pm\Theta^\pm\bar\Theta^\mp\mathcal{P}_\mp
\mp\frac{2}{Q^2}\partial\bar\Theta^\pm,
\nonumber\\
%%%%%%%%%%%%%%%%%
  \bar Q^\pm=&\bar{\mathcal{P}}_\mp.\label{slcurrent}
\end{align}
These currents satisfy the affine Lie superalgebra $sl(1|2)^{(1)}$ 
at level $k$:
\begin{alignat}{2}
 &J^{++}(z)J^{--}(w)\sim\frac{k}{(z-w)^2}-\frac{2J^3(w)}{z-w},&
%%%%%
 &J^3(z)J^{\pm\pm}(w) \sim \frac{\pm J^{\pm\pm}(w)}{z-w},\nonumber\\
%%%%%%%%%%%%%%%%%%%%%
 &J^3(z)J^3(w)\sim\frac{-k/2}{(z-w)^2},&
%%%%%
 &J^Y(z)J^Y(w)\sim \frac{2(k-2)}{(z-w)^2},\nonumber\\
%%%%%%%%%%%%%%%%%%%%%
 &J^{++}(z)Q^-(w)\sim-\frac{Q^+(w)}{z-w},&
%%%%%
 &J^{++}(z)\bar Q^-(w)\sim-\frac{\bar Q^+(w)}{z-w},\nonumber\\
%%%%%%%%%%%%%%%%%%%%%
 &J^{--}(z)Q^+(w)\sim\frac{Q^-(w)}{z-w},&
%%%%%
 &J^{--}(z)\bar Q^+(w)\sim\frac{\bar Q^-(w)}{z-w},\nonumber\\
%%%%%%%%%%%%%%%%%%%%%
 &J^3(z) Q^\pm(w)\sim\pm\frac{1}{2}\frac{ Q^\pm(w)}{z-w},&
%%%%%
 &J^3(z)\bar Q^\pm\sim\pm\frac{1}{2}\frac{\bar Q^\pm(w)}{z-w},\nonumber\\
%%%%%%%%%%%%%%%%%%%%%
 &J^Y(z) Q^\pm(w)\sim-\frac{ Q^\pm(w)}{z-w},&
%%%%%
 &J^Y(z)\bar Q^\pm(w)\sim\frac{\bar Q^\pm(w)}{z-w},\nonumber\\
%%%%%%%%%%%%%%%%%%%%%
& Q^+(z)\bar Q^+(w)\sim-\frac{J^{++}(w)}{z-w},&
%%%%%
& Q^-(z)\bar Q^-(w)\sim-\frac{J^{--}(w)}{z-w},\nonumber\\
%%%%%%%%%%%%%%%%%%%%%
& \bar Q^+(z) Q^-(w)\sim\frac{k}{(z-w)^2}
-\frac{(J^3-\frac{1}{2}J^Y)(w)}{z-w},& &\nonumber\\
%%%%%%%%%%%%%%%%%%%%%
& \bar Q^-(z) Q^+(w)\sim\frac{k}{(z-w)^2}
-\frac{(J^3+\frac{1}{2}J^Y)(w)}{z-w}. & &\label{salgebra}
\end{alignat}

The $N=2$ superconformal generators can be written 
in terms of these currents as\cite{Ito}
\begin{align}
 \mathcal{L}_n=&\oint\dz {}^\otimes_\otimes\Bigg(
-\frac{1}{2}n(n+1)\gamma^{n-1}J^{++}+(n^2-1)\gamma^nJ^3
-\frac{1}{2}n(n-1)\gamma^{n+1}J^{--}
\Bigg){}^\otimes_\otimes,\nonumber\\ 
%%%%%%%%%%%%%%%%%%%
\mathcal{G}^+_r=&\oint\dz{}^\otimes_\otimes\Bigg(
\left(r+\frac{1}{2}\right)\gamma^{r-\frac{1}{2}}\bar Q^+
-\left(r-\frac{1}{2}\right)\gamma^{r+\frac{1}{2}}\bar Q^-
\Bigg){}^\otimes_\otimes,\nonumber\\
%%%%%%%%%%%%%%%%%%%
\mathcal{G}^-_r=&\oint\dz{}^\otimes_\otimes\Bigg(
\left(r+\frac{1}{2}\right)\gamma^{r-\frac{1}{2}}Q^+
-\left(r-\frac{1}{2}\right)\gamma^{r+\frac{1}{2}}Q^-
\nonumber\\
&\hspace{30mm}
-\left(r^2-\frac{1}{4}\right)(\bar\Theta^++\gamma\bar\Theta^-)
\gamma^{r-\frac{3}{2}}(J^{++}-2\gamma J^3+\gamma^2J^{--})
\Bigg){}^\otimes_\otimes,\nonumber\\
%%%%%%%%%%%%%%%%%%%
\mathcal{I}_n=&\oint\dz{}^\otimes_\otimes\Bigg(
\gamma^nJ^Y+n\gamma^{n-1}\Big(
(\Theta^++\gamma\Theta^-)(\mathcal{P}_--\gamma\mathcal{P}_+)
\nonumber\\
&\hspace{60mm}
-(\bar\Theta^++\gamma\bar\Theta^-)(\bar{\mathcal{P}}_--\gamma\bar{\mathcal{P}}_+)
\Big)\Bigg){}^\otimes_\otimes,
\label{scg}
\end{align}
where $\onorm{\ }$ denotes normal ordering with respect to 
the coefficient fields and the $sl(1|2)$ {\it currents}.

\section{The Similarity Transformation to Real Variables}\label{appB}

The hybrid fields given in the text are analogs of chiral coordinates.
They are convenient for actual calculations but they possess some nontrivial 
hermiticity properties.
We can obtain the corresponding {\it real} variables with proper hermiticity
by carrying out the similarity transformation generated by
\begin{align}
  R=&\oint\dz\Bigg(
\frac{1}{2}\Theta^+\bar\Theta^+e^{\beta i( X^0+ X^1)}
\left(\frac{1}{\beta}i\partial X^1+\frac{1}{Q}\partial\phi_L\right)\nonumber\\
&\hspace{1cm}
-\frac{1}{2}\Theta^+\bar\Theta^-\left(\frac{1}{\beta}i\partial X^0
-\frac{1}{Q}i\partial\widehat Y+i\partial\rho
-\Theta^-\mathcal{P}_-+\bar\Theta^+\bar{\mathcal{P}}_+\right)\nonumber\\
&\hspace{1cm}
-\frac{1}{2}\Theta^-\bar\Theta^+\left(\frac{1}{\beta}i\partial X^0
+\frac{1}{Q}i\partial\widehat Y-i\partial\rho
+\Theta^+\mathcal{P}_+-\bar\Theta^-\bar{\mathcal{P}}_-\right)\nonumber\\
&\hspace{1cm}
+\frac{1}{2}\Theta^-\bar\Theta^-e^{-\beta i( X^0+ X^1)}
\left(\frac{1}{\beta}i\partial X^1-\frac{1}{Q}\partial\phi_L\right)\nonumber\\
&\hspace{1cm}
+\frac{1}{2}\Theta^+\Theta^-\bar\Theta^+\bar\Theta^-\left(
\frac{1}{Q}i\partial\widehat Y-i\partial\rho\right)
+\frac{1}{2}\left(\frac{1}{Q^2}-1\right)\Theta^+\Theta^-
\partial\left(\bar\Theta^+\bar\Theta^-\right)
\Bigg).
\end{align}
Under this transformation, $\mathcal{L}_n$ and $\mathcal{I}_n$ are
invariant, but the supercharges $\mathcal{G}^\pm_r$ are changed into 
the symmetric forms with proper hermiticity:
\begin{align}
\widehat{\mathcal{G}}^+_r=&\oint\dz\Bigg(
\left(r+\frac{1}{2}\right)\gamma^{r-\frac{1}{2}}\bigg(\bar{\mathcal{P}}_-
+\frac{1}{2}\Theta^-\gamma\left(\frac{1}{\beta}i\partial X^1
-\frac{1}{Q}\partial\phi_L\right)
\nonumber\\
&\hspace{2cm}
-\frac{1}{2}\Theta^+
\left(\frac{1}{\beta}i\partial X^0-\frac{1}{Q}i\partial\widehat Y
+i\partial\rho-\Theta^-\mathcal{P}_-
-\left(r-\frac{3}{2}\right)\bar\Theta^+\bar{\mathcal{P}}_+
+\left(r-\frac{1}{2}\right)\bar\Theta^-\bar{\mathcal{P}}_-\right)
\nonumber\\
&\hspace{2cm}
-\frac{1}{2}\left(r+\frac{1}{2}\right)\Theta^-\bar\Theta^+\bar{\mathcal{P}}_-
+\frac{1}{4}\Theta^+\Theta^-\bar\Theta^+
\left(\frac{1}{Q}i\partial\widehat Y-i\partial\rho\right)
\nonumber\\
&\hspace{2cm}
-\frac{1}{4}\left(\frac{1}{Q^2}-1\right)\partial(\Theta^+\Theta^-)\bar\Theta^+
+\frac{1}{2}\partial\Theta^+
\bigg)
\nonumber\\
&\hspace{1cm}
-\left(r-\frac{1}{2}\right)\gamma^{r+\frac{1}{2}}\bigg(\bar{\mathcal{P}}_+
+\frac{1}{2}\Theta^+\gamma^{-1}\left(\frac{1}{\beta}i\partial X^1
+\frac{1}{Q}\partial\phi_L\right)
\nonumber\\
&\hspace{2cm}
-\frac{1}{2}\Theta^-\left(\frac{1}{\beta}i\partial X^0
+\frac{1}{Q}i\partial\widehat Y-i\partial\rho
+\Theta^+\mathcal{P}_+
+\left(r+\frac{1}{2}\right)\bar\Theta^+\bar{\mathcal{P}}_+
-\left(r+\frac{3}{2}\right)\bar\Theta^-\bar{\mathcal{P}}_-\right)
\nonumber\\
&\hspace{2cm}
-\frac{1}{2}\left(r-\frac{1}{2}\right)\Theta^+\bar\Theta^-\bar{\mathcal{P}}_+
-\frac{1}{4}\Theta^+\Theta^-\bar\Theta^-
\left(\frac{1}{Q}i\partial\widehat Y-i\partial\rho\right)
\nonumber\\
&\hspace{2cm}
+\frac{1}{4}\left(\frac{1}{Q^2}-1\right)\partial(\Theta^+\Theta^-)\bar\Theta^-
-\frac{1}{2}\partial\Theta^-\bigg)
\nonumber\\
&\hspace{1cm}
-\frac{1}{2}\left(r^2-\frac{1}{4}\right)\left(
\gamma^{r-\frac{3}{2}}\Theta^+\bar\Theta^+\bar{\mathcal{P}}_-
-\gamma^{r+\frac{3}{2}}\Theta^-\bar\Theta^-\bar{\mathcal{P}}_+\right)
\Bigg),\nonumber\\
%%%%%%%%%%%%%%%%%%%%%%%%
=&\oint\dz{}^\times_\times\Bigg(
\left(r+\frac{1}{2}\right)\gamma^{r-\frac{1}{2}}\hat{\bar Q}^+
-\left(r-\frac{1}{2}\right)\gamma^{r+\frac{1}{2}}\hat{\bar Q}^-
\nonumber\\
&\hspace{3cm}
-\frac{1}{2}\left(r^2-\frac{1}{4}\right)\gamma^{r-\frac{3}{2}}
\left(\Theta^++\gamma\Theta^-\right)
\left(J^{++}-2\gamma J^3+\gamma^2J^{--}\right)
\Bigg){}^\times_\times,\nonumber\\
%%%%%%%%%%%%%%%%%%%
\widehat{\mathcal{G}}^-_r=&\oint\dz\Bigg(
\left(r+\frac{1}{2}\right)\gamma^{r-\frac{1}{2}}\bigg(\mathcal{P}_-
+\frac{1}{2}\bar\Theta^-\gamma\left(\frac{1}{\beta}i\partial X^1
-\frac{1}{Q}\partial\phi_L\right)
\nonumber\\
&\hspace{2cm}
-\frac{1}{2}\bar\Theta^+\left(\frac{1}{\beta}i\partial X^0+
\frac{1}{Q}i\partial\widehat Y-i\partial\rho
-\left(r-\frac{3}{2}\right)\Theta^+\mathcal{P}_+
+\left(r-\frac{1}{2}\right)\Theta^-\mathcal{P}_-
-\bar\Theta^-\bar{\mathcal{P}}_-\right)
\nonumber\\
&\hspace{2cm}
+\frac{1}{2}\left(r+\frac{1}{2}\right)\Theta^+\bar\Theta^-\mathcal{P}_-
-\frac{1}{4}\Theta^+\bar\Theta^+\bar\Theta^-
\left(\frac{1}{Q}i\partial\widehat Y-i\partial\rho\right)
\nonumber\\
&\hspace{2cm}
-\frac{1}{4}\left(\frac{1}{Q^2}-1\right)\Theta^+\partial(\bar\Theta^+\bar\Theta^-)
+\frac{1}{2}\partial\bar\Theta^+
\bigg)
\nonumber\\
&\hspace{1cm}
-\left(r-\frac{1}{2}\right)\gamma^{r+\frac{1}{2}}\bigg(\mathcal{P}_+
+\frac{1}{2}\bar\Theta^+\gamma^{-1}\left(\frac{1}{\beta}i\partial X^1
+\frac{1}{Q}\partial\phi_L\right)
\nonumber\\
&\hspace{2cm}
-\frac{1}{2}\bar\Theta^-\left(\frac{1}{\beta}i\partial X^0
-\frac{1}{Q}i\partial\widehat Y+i\partial\rho
+\left(r+\frac{1}{2}\right)\Theta^+\mathcal{P}_+
-\left(r+\frac{3}{2}\right)\Theta^-\mathcal{P}_-
+\bar\Theta^+\bar{\mathcal{P}}_+\right)
\nonumber\\
&\hspace{2cm}
+\frac{1}{2}\left(r-\frac{1}{2}\right)\Theta^-\bar\Theta^+\mathcal{P}_+
+\frac{1}{4}\Theta^-\bar\Theta^+\bar\Theta^-
\left(\frac{1}{Q}i\partial\widehat Y-i\partial\rho\right)
\nonumber\\
&\hspace{2cm}
+\frac{1}{4}\left(\frac{1}{Q^2}-1\right)\Theta^-\partial(\bar\Theta^+\bar\Theta^-)
-\frac{1}{2}\partial\bar\Theta^-\bigg)
\nonumber\\
&\hspace{1cm}
+\frac{1}{2}\left(r^2-\frac{1}{4}\right)\left(
\gamma^{r-\frac{3}{2}}\Theta^+\bar\Theta^+\mathcal{P}_-
-\gamma^{r+\frac{3}{2}}\Theta^-\bar\Theta^-\mathcal{P}_+\right)
\Bigg),\nonumber\\
%%%%%%%%%%%%%%%%%%%
=&\oint\dz{}^\times_\times\Bigg(
\left(r+\frac{1}{2}\right)\gamma^{r-\frac{1}{2}}\hat Q^+
-\left(r-\frac{1}{2}\right)\gamma^{r+\frac{1}{2}}\hat Q^-
\nonumber\\
&\hspace{3cm}
-\frac{1}{2}\left(r^2-\frac{1}{4}\right)\gamma^{r-\frac{3}{2}}
\left(\bar\Theta^++\gamma\bar\Theta^-\right)
\left(J^{++}-2\gamma J^3+\gamma^2J^{--}\right)
\Bigg){}^\times_\times.\label{scharge2}
\end{align}
Here, the transformed fermionic currents
$(\hat{Q}^\pm,\hat{\bar Q}^\pm)$
also have symmetric forms:
\begin{align}
 \hat Q^\pm=&\mathcal{P}_\mp
+\frac{1}{2}\bar\Theta^\mp e^{\mp\beta i( X^0+ X^1)}
\left(\frac{1}{\beta}i\partial X^1\mp
\frac{1}{Q}\partial\phi_L\right)\nonumber\\
&\hspace{5mm}
-\frac{1}{2}\bar\Theta^\pm\left(
\frac{1}{\beta}i\partial X^0\pm\frac{1}{Q}i\partial\widehat Y
\mp i\partial\rho\pm\Theta^\pm\mathcal{P}_\pm
\mp\bar\Theta^\mp\bar{\mathcal{P}}_\mp\right)
\nonumber\\
&\hspace{5mm}
\pm\frac{1}{2}\Theta^\pm\bar\Theta^\mp\mathcal{P}_\mp
-\frac{1}{4}\Theta^\pm\bar\Theta^\pm\bar\Theta^\mp
\left(\frac{1}{Q}i\partial\widehat Y-i\partial\rho\right)
\nonumber\\
&\hspace{5mm}
-\frac{1}{4}\left(\frac{1}{Q^2}-1\right)\Theta^\pm
\partial\left(\bar\Theta^\pm\bar\Theta^\mp\right)
\mp\frac{1}{Q^2}\partial\bar\Theta^\pm,\nonumber\\
%%%%%%%%%%%%%%%%%%%%%%%%%%%%%%%%%%%%%%%%%%
  \hat{\bar Q}^\pm=&\bar{\mathcal{P}}_\mp
+\frac{1}{2}\Theta^\mp e^{\mp\beta i( X^0+ X^1)}
\left(\frac{1}{\beta}i\partial X^1\mp
\frac{1}{Q}\partial\phi_L\right)\nonumber\\
&\hspace{5mm}
-\frac{1}{2}\Theta^\pm\left(
\frac{1}{\beta}i\partial X^0\mp\frac{1}{Q}i\partial\widehat Y
\pm i\partial\rho\mp\Theta^\mp\mathcal{P}_\mp
\pm\bar\Theta^\pm\bar{\mathcal{P}}_\pm\right)
\nonumber\\
&\hspace{5mm}
\mp\frac{1}{2}\Theta^\mp\bar\Theta^\pm\bar{\mathcal{P}}_\mp
+\frac{1}{4}\Theta^\pm\Theta^\mp\bar\Theta^\pm
\left(\frac{1}{Q}i\partial\widehat Y-i\partial\rho\right)
\nonumber\\
&\hspace{5mm}
-\frac{1}{4}\left(\frac{1}{Q^2}-1\right)
\partial\left(\Theta^\pm\Theta^\mp\right)\bar\Theta^\pm
\mp\frac{1}{Q^2}\partial\Theta^\pm.
\end{align}
We note that the new currents $(J^{\pm\pm},J^3,J^Y,
\hat{Q}^\pm,\hat{\bar Q}^\pm)$ satisfy the same
$sl(1|2)$ current superalgebra, (\ref{salgebra}).\footnote{
This can be easily seen from the fact that the original and the new
currents are related by the similarity transformation generated 
by $R+R'$, where $R'=\frac{1}{Q^2}\oint\dz 
(\Theta^+\partial\bar\Theta^--\Theta^-\partial\bar\Theta^+)$.}

The similarity transformation also acts on the world-sheet 
$N=2$ superconformal generators. The bosonic generators $T$ and $I$
are invariant, but the fermionic $G^\pm$ are transformed as
\begin{align}
G^+=&\frac{Q}{\sqrt{2}}{}^\times_\times e^{-i\rho}{\cal T}\hat{\bar d}_+
\hat{\bar d}_-{}^\times_\times +\widehat G^+_{\mathcal{N}/U(1)},\nonumber\\
%%%%%%%%%%%%%%%
G^-=&\frac{Q}{\sqrt{2}}{}^\times_\times e^{i\rho}{\cal T}\hat d_+
\hat d_-{}^\times_\times +\widehat G^-_{\mathcal{N}/U(1)},
\label{wcharge2}
\end{align}
where
\begin{align}
  {\cal T}=&1
+\frac{1}{2}\Theta^+\bar\Theta^-
-\frac{1}{2}\Theta^-\Theta^-
+\frac{1}{4}\Theta^+\Theta^-\bar\Theta^+\bar\Theta^-,\nonumber\\
%%%%%%%%%%%%%%%%%%%%%%
\hat d_\mp=&\mathcal{P}_\mp
-\frac{1}{2}\bar\Theta^\mp e^{\mp\beta i( X^0+ X^1)}
\left(\frac{1}{\beta}i\partial X^1\mp
\frac{1}{Q}\partial\phi_L\right)
\nonumber\\
&
+\frac{1}{2}\bar\Theta^\pm\left(
\frac{1}{\beta}i\partial X^0\pm\frac{1}{Q}i\partial\widehat Y
\mp i\partial\rho
\pm\Theta^\pm\mathcal{P}_\pm\mp\bar\Theta^\mp\bar{\mathcal{P}}_\mp\right)
\nonumber\\
&
\mp\frac{1}{2}\Theta^\pm\bar\Theta^\mp\mathcal{P}_\mp
+\frac{3}{4}\Theta^\pm\bar\Theta^\pm\bar\Theta^\mp
\left(\frac{1}{Q}i\partial\widehat Y-i\partial\rho\right)
\nonumber\\
&
+\frac{3}{4}\left(\frac{1}{Q^2}-1\right)\Theta^\pm
\partial(\bar\Theta^\pm\bar\Theta^\mp)
-\frac{1}{2Q^2}\partial(\Theta^\pm\bar\Theta^\pm\bar\Theta^\mp),
\nonumber\\
%%%%%%%%%%%%%%%%%%%%%%
 \hat{\bar d}_\pm=&\bar{\mathcal{P}}_\pm
-\frac{1}{2}\Theta^\pm e^{\pm\beta i( X^0+ X^1)}
\left(\frac{1}{\beta}i\partial X^1\pm
\frac{1}{Q}\partial\phi_L\right)
\nonumber\\
&
+\frac{1}{2}\Theta^\mp\left(
\frac{1}{\beta}i\partial X^0\pm\frac{1}{Q}i\partial\widehat Y
\mp i\partial\rho
\pm\Theta^\pm\mathcal{P}_\pm\mp\bar\Theta^\mp\bar{\mathcal{P}}_\mp\right)
\nonumber\\
&
\mp\frac{1}{2}\Theta^\pm\bar\Theta^\mp\bar{\mathcal{P}}_\pm
+\frac{3}{4}\Theta^\pm\Theta^\mp\bar\Theta^\mp
\left(\frac{1}{Q}i\partial\widehat Y-i\partial\rho\right)
\nonumber\\
&
-\frac{3}{4}\left(\frac{1}{Q^2}-1\right)\bar\Theta^\mp
\partial(\Theta^\pm\Theta^\mp)
+\frac{1}{2Q^2}\partial(\Theta^\pm\Theta^\mp\bar\Theta^\mp).
\end{align}
The normal ordering denoted by $\norm{}$ in (\ref{scharge2}) and (\ref{wcharge2})
is that with respect to the new currents.

\section{Bosonic Transformation Laws of the Component Fields}\label{appC}

In this appendix, we present the bosonic transformation laws
$(\delta^\mathcal{L}_n,\delta^\mathcal{I}_n)$ of the component 
fields in the WZ-like gauge.

For $l=-1$, the bosonic transformations on the chiral
supermultiplet $(\varphi,d;\chi^\pm)$ are given by
\begin{align}
\delta^\mathcal{L}_n\varphi(m)=&
-\left(m+n(j-1)\right)\varphi(m-n),\nonumber\\
%%%%%%%%%%%%%%%%%%%%%%%
\delta^\mathcal{L}_nd(m-1)=&
-\left(m-1+n(j-1)\right)d(m-n-1),\nonumber\\
%%%%%%%%%%%%%%%%%%%%%%%
\delta^\mathcal{L}_n\chi^+(m)=&
-\left(m+n(j-1)+\frac{1}{2}(n^2-1)\right)\chi^+(m-n)
-\frac{n}{2}(n-1)\chi^-(m-n-1),\nonumber\\
%%%%%%%%%%%%%%%%%%%%%%%
\delta^\mathcal{L}_n\chi^-(m-1)=&
-\left(m-1+n(j-1)-\frac{1}{2}(n^2-1)\right)\chi^-(m-n-1)
+\frac{n}{2}(n+1)\chi^+(m-n),\nonumber\\
%%%%%%%%%%%%%%%%%%%%%%%%%%%%%%%%%%%%%%%%%%%%%%%%
\delta^\mathcal{I}_n\varphi(m)=&2q\varphi(m-n),\nonumber\\
%%%%%%%%%%%%%%%%%%%%%%%
\delta^\mathcal{I}_nd(m-1)=&(2d+2)d(m-n-1),\nonumber\\
%%%%%%%%%%%%%%%%%%%%%%%
\delta^\mathcal{I}_n\chi^+(m)=&
\left(2q+n+1\right)\chi^+(m-n)+n\chi^-(m-n-1),\nonumber\\
%%%%%%%%%%%%%%%%%%%%%%%
\delta^\mathcal{I}_n\chi^-(m-1)=&
\left(2q-n+1\right)\chi^-(m-n-1)-n\chi^+(m-n).
\end{align}

For $l=0$ and $Q_\mathcal{N}\ne0$,
the off-shell fields are the massive vector supermultiplet
$\ (v_{\pm\pm},\ v_3,\\
v_Y,\varphi,d;\psi_\pm,\chi^\pm,\bar{\chi}^\pm)$,
and their transformations are given by
\begin{align}
%%%%%%%%%%%%%%%%%%%%%% 
\delta^\mathcal{L}_nv_{++}(m-1)=&
-\left(m-1+n(j-1)-n^2+1\right)v_{++}(m-n-1)-n(n+1)v_3(m-n),\nonumber\\
%%%%%%%%%%%%%%%%%%%%%% 
\delta^\mathcal{L}_nv_3(m)=&
-\left(m+n(j-1)\right)v_3(m-n)\nonumber\\
&
+\frac{1}{2}n(n-1)v_{++}(m-n-1)
-\frac{1}{2}n(n+1)v_{--}(m-n+1),\nonumber\\
%%%%%%%%%%%%%%%%%%%%%% 
\delta^\mathcal{L}_nv_{--}(m+1)=&
-\left(m+1+n(j-1)+n^2-1\right)v_{--}(m-n+1)+n(n-1)v_3(m-n),\nonumber\\
%%%%%%%%%%%%%%%%%%%%%% 
\delta^\mathcal{L}_nv_Y(m)=&
-\left(m+n(j-1)\right)v_Y(m-n),\nonumber\\
%%%%%%%%%%%%%%%%%%%%%% 
\delta^\mathcal{L}_n\varphi(m+1)=&
-\left(m+1+n(j-1)\right)\varphi(m-n+1),\nonumber\\
%%%%%%%%%%%%%%%%%%%%%% 
\delta^\mathcal{L}_nd(m)=&
-\left(m+n(j-1)\right)d(m-n),\nonumber\\
%%%%%%%%%%%%%%%%%%%%%% 
\delta^\mathcal{L}_n\psi_+(m)=&
-\left(m+n(j-1)-\frac{1}{2}(n^2-1)\right)\psi_+(m-n)\nonumber\\
&
-\frac{1}{2}n(n+1)\psi_-(m-n+1),\nonumber\\
%%%%%%%%%%%%%%%%%%%%%% 
\delta^\mathcal{L}_n\psi_-(m+1)=&
-\left(m+1+n(j-1)+\frac{1}{2}(n^2-1)\right)\psi_-(m-n+1)\nonumber\\
&
+\frac{1}{2}n(n-1)\psi_+(m-n),\nonumber\\
%%%%%%%%%%%%%%%%%%%%%% 
\delta^\mathcal{L}_n\chi^+(m+1)=&
-\left(m+1+n(j-1)+\frac{1}{2}(n^2-1)\right)\chi^+(m-n+1)\nonumber\\
&
-\frac{1}{2}n(n-1)\chi^-(m-n),\nonumber\\
%%%%%%%%%%%%%%%%%%%%%% 
\delta^\mathcal{L}_n\chi^-(m)=&
-\left(m+n(j-1)-\frac{1}{2}(n^2-1)\right)\chi^-(m-n)\nonumber\\
&
+\frac{1}{2}n(n+1)\chi^+(m-n+1),\nonumber\\
%%%%%%%%%%%%%%%%%%%%%% 
\delta^\mathcal{L}_n\bar\chi^+(m)=&
-\left(m+n(j-1)+\frac{1}{2}(n^2-1)\right)\bar\chi^+(m-n)\nonumber\\
&
-\frac{1}{2}n(n-1)\bar\chi^-(m-n-1),\nonumber\\
%%%%%%%%%%%%%%%%%%%%%% 
\delta^\mathcal{L}_n\bar\chi^-(m-1)=&
-\left(m-1+n(j-1)-\frac{1}{2}(n^2-1)\right)\bar\chi^-(m-n-1)\nonumber\\
&
+\frac{1}{2}n(n+1)\bar\chi^+(m-n),\nonumber\\
%%%%%%%%%%%%%%%%%%%%%% 
\delta^\mathcal{I}_nv_{++}(m-1)=&
2qv_{++}(m-n-1)+2nv_Y(m-n),\nonumber\\
%%%%%%%%%%%%%%%%%%%%%% 
\delta^\mathcal{I}_nv_3(m)=&
2qv_3(m-n)+2nv_Y(m-n),\nonumber\\
%%%%%%%%%%%%%%%%%%%%%% 
\delta^\mathcal{I}_nv_{--}(m+1)=&
2qv_{--}(m-n+1)+2nv_Y(m-n),\nonumber\\
%%%%%%%%%%%%%%%%%%%%%% 
\delta^\mathcal{I}_nv_Y(m)=&
2qv_Y(m-n)+2nv_3(m-n)-nv_{++}(m-n-1)-nv_{--}(m-n+1),\nonumber\\
%%%%%%%%%%%%%%%%%%%%%% 
\delta^\mathcal{I}_n\varphi(m+1)=&
2(q-1)\varphi(m-n+1),\nonumber\\
%%%%%%%%%%%%%%%%%%%%%% 
\delta^\mathcal{I}_nd(m)=&
2qd(m-n),\nonumber\\
%%%%%%%%%%%%%%%%%%%%%% 
\delta^\mathcal{I}_n\psi_+(m)=&
(2q+n-1)\psi_+(m-n)-n\psi_-(m-n+1),\nonumber\\
%%%%%%%%%%%%%%%%%%%%%% 
\delta^\mathcal{I}_n\psi_-(m+1)=&
(2q-n-1)\psi_-(m-n+1)+n\psi_+(m-n),\nonumber\\
%%%%%%%%%%%%%%%%%%%%%% 
\delta^\mathcal{I}_n\chi^+(m+1)=&
(2q+n-1)\chi^+(m-n+1)+n\chi^-(m-n),\nonumber\\
%%%%%%%%%%%%%%%%%%%%%% 
\delta^\mathcal{I}_n\chi^-(m)=&
(2q-n-1)\chi^-(m-n)-n\chi^+(m-n+1),\nonumber\\
%%%%%%%%%%%%%%%%%%%%%% 
\delta^\mathcal{I}_n\bar\chi^+(m)=&
(2q+n+1)\bar\chi^+(m-n)+n\chi^-(m-n-1),\nonumber\\
%%%%%%%%%%%%%%%%%%%%%% 
\delta^\mathcal{I}_n\bar\chi^-(m-1)=&
(2q+n+1)\bar\chi^-(m-n-1)+n\bar\chi^+(m-n). 
\end{align}
We can easily confirm that the supersymmetry transformations (\ref{susytf1})
and (\ref{susytf2}) actually satisfy the algebra (\ref{sca}) 
by using these bosonic transformation laws.

For the compactification independent case, $l=0$ and $Q_\mathcal{N}=0$,
the transformations on the massless vector supermultiplet
$(v_{\pm\pm},v_3,v_Y,d;\chi^\pm,\bar{\chi}\pm)$ are obtained as
\begin{align}
%%%%%%%%%%%%%%%%%%%%%
\delta^\mathcal{L}_nv_{++}(m-1)=&-(m-1+n(j-1)-n^2+1)v_{++}(m-n-1)
-n(n+1)v_3(m-n),\nonumber\\
%%%%%%%%%%%%%%%%%%%%%
\delta^\mathcal{L}_nv_3(m)=&-(m+n(j-1))v_3(m-n)\nonumber\\
&
+\frac{1}{2}n(n-1)v_{++}(m-n-1)-\frac{1}{2}n(n+1)v_{--}(m-n+1),\nonumber\\
%%%%%%%%%%%%%%%%%%%%%
\delta^\mathcal{L}_nv_{--}(m+1)=&-(m+n(j-1)+n^2-1)v_{--}(m-n+1)
+n(n-1)v_3(m-n),\nonumber\\
%%%%%%%%%%%%%%%%%%%%%
\delta^\mathcal{L}_nv_Y(m)=&-(m+n(j-1))v_Y(m-n),\nonumber\\
%%%%%%%%%%%%%%%%%%%%%
\delta^\mathcal{L}_nd(m)=&-(m+n(j-1))d(m-n),\nonumber\\
%%%%%%%%%%%%%%%%%%%%%
\delta^\mathcal{L}_n\chi^+(m+1)=&-(m+1+n(j-1)+\frac{1}{2}(n^2-1))\chi^+(m-n+1)
-\frac{1}{2}n(n-1)\chi^-(m-n),\nonumber\\
%%%%%%%%%%%%%%%%%%%%%
\delta^\mathcal{L}_n\chi^-(m)=&-(m+n(j-1)-\frac{1}{2}(n^2-1))\chi^-(m-n)
+\frac{1}{2}n(n+1)\chi^+(m-n+1),\nonumber\\
%%%%%%%%%%%%%%%%%%%%%
\delta^\mathcal{L}_n\bar\chi^+(m)=&-(m+n(j-1)+\frac{1}{2}(n^2-1))\bar\chi^+(m-n)
-\frac{1}{2}n(n-1)\bar\chi^-(m-n-1),\nonumber\\
%%%%%%%%%%%%%%%%%%%%%
\delta^\mathcal{L}_n\bar\chi^-(m-1)=&
-(m-1+n(j-1)-\frac{1}{2}(n^2-1))\bar\chi^-(m-n-1)
+\frac{1}{2}n(n+1)\bar\chi^+(m-n),\nonumber\\
%%%%%%%%%%%%%%%%%%%%%
\delta^\mathcal{I}_nv_{++}(m-1)=&2nv_Y(m-n),\nonumber\\
%%%%%%%%%%%%%%%%%%%%%
\delta^\mathcal{I}_nv_3(m)=&2nv_Y(m-n),\nonumber\\
%%%%%%%%%%%%%%%%%%%%%
\delta^\mathcal{I}_nv_{--}(m+1)=&2nv_Y(m-n),\nonumber\\
%%%%%%%%%%%%%%%%%%%%%
\delta^\mathcal{I}_nv_Y(m)=&2nv_3(m-n)
-nv_{++}(m-n-1)+nv_{--}(m-n+1),\nonumber\\
%%%%%%%%%%%%%%%%%%%%%
\delta^\mathcal{I}_nd(m)=&0,\nonumber\\
%%%%%%%%%%%%%%%%%%%%%
\delta^\mathcal{I}_n\chi^+(m+1)=&(n-1)\chi^+(m-n+1)+n\chi^-(m-n),\nonumber\\
%%%%%%%%%%%%%%%%%%%%%
\delta^\mathcal{I}_n\chi^-(m)=&-(n+1)\chi^-(m-n)-n\chi^+(m-n+1),\nonumber\\
%%%%%%%%%%%%%%%%%%%%%
\delta^\mathcal{I}_n\bar\chi^+(m)=&-(n-1)\bar\chi^+(m-n)-n\bar\chi^-(m-n-1),\nonumber\\
%%%%%%%%%%%%%%%%%%%%%
\delta^\mathcal{I}_n\bar\chi^-(m-1)=&(n+1)\bar\chi^-(m-n-1)+n\bar\chi^+(m-n).
\end{align}
We can confirm that the algebra (\ref{msusy}) holds for 
the transformations (\ref{susytf3}).

\end{document}